\documentclass[12pt,a4paper]{article}

\usepackage{amsmath,amssymb,graphicx}
\usepackage{xspace}
\usepackage{multirow}
\usepackage{cite}
\usepackage{lineno}

\newlength{\abstwidth}
\newcommand{\forcenewcommand}[1]{\providecommand{#1}{}\renewcommand{#1}}
\newcommand{\defparticle}[1]{
  \expandafter\forcenewcommand\csname #1\endcsname{{\mathrm{#1}}}
  \expandafter\forcenewcommand\csname #1bar\endcsname{{\bar{\mathrm{#1}}}}
}

\defparticle{g}
\defparticle{q}
\defparticle{u}
\defparticle{d}
\defparticle{s}
\defparticle{c}
\defparticle{b}
\defparticle{t}
\defparticle{e}
\defparticle{p}
\defparticle{n}
\defparticle{A}
\defparticle{B}
\defparticle{D}
\defparticle{K}
\defparticle{N}
\defparticle{Pb}
\defparticle{W}
\defparticle{Z}

\newcommand{\gammat}{\Gamma_{\mathrm{t}}}
\newcommand{\gammatG}{\Gamma_{\mathrm{t,G}}}
\newcommand{\gammatBW}{\Gamma_{\mathrm{t,BW}}}
\newcommand{\betat}{\beta_{\mathrm{t}}}
\newcommand{\mt}{m_{\mathrm{t}}}
\newcommand{\mtone}{m_{\mathrm{t} 1}}
\newcommand{\mttwo}{m_{\mathrm{t} 2}}
\newcommand{\mtbar}{\overline{m}_{\mathrm{t}}}
\newcommand{\as}{\alpha_{\mathrm{s}}}
\newcommand{\aem}{\alpha_{\mathrm{em}}}
\newcommand{\pT}{p_{\perp}}
\newcommand{\mhat}{\widehat{m}}
\newcommand{\shat}{\hat{s}}
\newcommand{\that}{\hat{t}}
\newcommand{\uhat}{\hat{u}}
\newcommand{\Ecm}{E_{\mathrm{CM}}}

\newcommand{\Pythia}{\textsc{Pythia}\xspace}

\newcommand{\eg}{\textit{e.g.}\xspace}
\newcommand{\ie}{\textit{i.e.}\xspace}
\newcommand{\cf}{\textit{cf.}\xspace}
\newcommand{\etal}{\textit{et~al.}\xspace}

\begin{document}

\sloppy
 
\pagestyle{empty}
 
\begin{flushright}
MCNET-26-12\\ 
TTK-26-16\\ 
P3H-26-039\\
May 2026
\end{flushright}

\vspace{\fill}

\begin{center}
{\Huge\bf Top Pair Threshold Revisited}\\[4mm]
{\Large Torbjörn Sjöstrand$^1$, Valery A. Khoze$^2$ and
Christian T. Preuss$^3$} \\[3mm]
{\it $^1$Department of Physics, Lund University\\[1mm]
$^2$Institute for Particle Physics Phenomenology,
University of Durham\\[1mm]
$^3$Institute for Theoretical Particle Physics and Cosmology,
RWTH Aachen University}\\[1mm]
\end{center}

\vspace{\fill}

\begin{center}
\begin{minipage}{\abstwidth}
{\bf Abstract}\\[2mm]
Recently the CMS and ATLAS collaborations have found evidence for
an unexpectedly large $\t\tbar$ cross section in the threshold 
region, with an excess of the order of 5--10~pb relative to 
continuum perturbative calculations. A convenient approach to the 
theoretical study of this region, unifying the above- and 
below-threshold behaviour, is the non-relativistic Green's 
function formalism. It was first applied to top production more than 
35 years ago, well before the discovery of the top. We therefore
revive and dissect the old formalism, and put it back together in a more
consistent form, suited for Monte Carlo event generation. Combined
with some practical prescriptions, it can be applied
to current conditions. As an example, the below-threshold cross
section comes out to be of the order of 6.5~pb, \ie comparable with
the CMS and ATLAS numbers. The new code is publicly available in the 
\textsc{Pythia} event generator so can be used for more detailed 
comparisons. 
\end{minipage}
\end{center}

\vspace{\fill}

\phantom{dummy}

\clearpage

\pagestyle{plain}
\setcounter{page}{1}

\section{Introduction}

The concept of a top quark gained traction with the Kobayashi--Maskawa 
observation in 1973 that a third generation offered a natural
mechanism for $CP$ violation \cite{Kobayashi:1973fv}. The subsequent 
discovery of the $\tau$ lepton in 1975 \cite{Perl:1975bf} and the 
$\Upsilon$ in 1977 \cite{E288:1977xhf} confirmed the existence of a 
third generation. It was generally guessed that the discovery of top 
would follow soon thereafter, at around a 15~GeV mass, based on the 
approximate factor-of-three jump between the $\s$, $\c$ and $\b$ quark 
masses. A spin-1 toponium state would predominantly decay to three 
gluons, as supported by 1978 PLUTO $\Upsilon$ studies
\cite{PLUTO:1978jrw}, and a spin-0 to two. Successive $\e^+\e^-$ 
colliders failed to find the top, pushing direct limits upwards. 
The 1987 ARGUS observation of $\B^0 - \Bbar^0$ oscillations 
\cite{ARGUS:1987xtv} indirectly suggested a more massive top, possibly 
above 100~GeV, see e.g.\cite{Ali:1987jv,Blinov:1988cc}, 
and later LEP electroweak results pointed in the same 
direction \cite{LEP:1991hsu}. Isolating top production against the 
sizeable background at hadron colliders is difficult, and led to some 
false starts before CDF found first evidence in 1994
\cite{CDF:1994vkk}, and CDF together with D0 announced discovery in 
1995 \cite{CDF:1995wbb,D0:1995jca}.

As the top mass limit went up, the expected toponium phenomenology
changed. For a top mass above 100~GeV the weak top decay width exceeds 
the toponium strong ditto. For $\mt > 125$~GeV the top quark decay time
also becomes smaller than the revolution time of a $\t\tbar$ state
(or the inverse of the binding energy),
challenging the concept of a bound state in a traditional sense 
\cite{Bigi:1986jk}. Nevertheless some estimates of toponium cross 
sections could be made. The large width smeared out the distinction 
between below-threshold pseudo-bound states and above-threshold 
Coulomb effects, however. The Green's function formalism of
non-relativistic QCD (NRQCD), which in a Feynman-graph language is
connected with the summation of the diagrams with an infinite number 
of uncrossed Coulomb gluon exchanges (Coulomb ladder), offers a 
unified approach to the whole threshold region. It was first
introduced for top at $\e^+\e^-$ colliders by Fadin and Khoze (FK) 
\cite{Fadin:1987wz,Fadin:1988fn}, and then extended to hadron colliders
\cite{Fadin:1989fd,Fadin:1990wx}, with some further information in 
\cite{Fadin:1991zw}. In \cite{Fadin:1987wz,Fadin:1988fn} the exact 
solution to the nonrelativistic Coulomb problem was used. 
In the subsequent years this formalism was further developed, 
\eg using more sophisticated QCD potentials, and applied both for 
$\e^+\e^-$ \cite{Kwong:1990iy,Strassler:1990nw,Hoang:2000yr} and 
hadron \cite{Hagiwara:2008df,Kiyo:2008bv,Sumino:2010bv,Ju:2020otc} 
collider phenomenology. Nevertheless, since then the emphasis
has been on comparisons between data and higher-order perturbative
calculations, as pioneered by Nason \etal \cite{Nason:1987xz}. 

But now considerable interest has been generated by the observation
of an excess, relative to such perturbative calculations, of
$\t\tbar$ production cross sections in the threshold region. 
Three analyses have been presented in the last two years.
The first, by CMS \cite{CMS:2025kzt}, reported results 
in the dilepton channel as $\sigma(\eta_{\t}) = 8.8^{+1.2}_{-1.4}$~pb,
with $\eta_{\t}$ a pseudoscalar ``toponium'' state. 
The second, by ATLAS \cite{ATLAS:2026dbe}, also in the dilepton
channel, gave $\sigma_{\mathrm{excess}} = 9.3^{+ 1.4}_{- 1.3}$~pb.  
The third, by CMS \cite{CMS-PAS-TOP-25-002}, now in the single lepton
channel, gave $\sigma_{\mathrm{excess}} = 5.1 \pm 0.9$~pb.  
In all three cases for 13~TeV $\p\p$ collisions, and corrected for 
branching ratios, efficiencies, backgrounds, etc.

One possible interpretation is that of below-threshold production
of pseudo-bound states. Unfortunately the experimental $\t\tbar$ mass
resolution of some tens of GeV is not good enough to separate below-
and above-threshold contributions, when the typical bound-state scale
is of the order of 2~GeV. On the theory side it is therefore relevant
to revive and further develop formalisms that provide a unified
description of the whole threshold region. 

The CMS analyses mainly rely on the simplified model of $\eta_{\t}$
production by Fuks \etal \cite{Fuks:2021xje} (see also 
\cite{Maltoni:2024tul}), implemented in \textsc{MadGraph5}\_aMC@NLO
\cite{Alwall:2014hca}. A more sophisticated option is provided 
by a later Fuks \etal article \cite{Fuks:2024yjj}, wherein a NRQCD 
Green's function is derived numerically and stored in data files
that can be used for reweighting purposes. Code is also provided  
how these tables can be used to generate events with 
\textsc{MadGraph5}\_aMC@NLO. This option is used by ATLAS and as an
alternative in the second CMS article. It has also been further studied
by the authors \cite{Fuks:2025wtq,Fuks:2025toq}. Although not a
generator, often results are also compared with the NRQCD calculations
of  Garzelli \etal \cite{Garzelli:2024uhe,Garzelli:2026ctb}. 
The NRQCD studies above are based on more sophisticated QCD potentials
than in FK, but this then leads to the need for numerical solutions
that offer reduced flexibility.

In addition to the ``signal'' software, a full analysis also needs
to simulate the ``background'' non-toponium-related processes.
Typically CMS and ATLAS combine several tools that provide for
NNLO QCD + NLO electroweak matrix elements, threshold resummation,
non-resonant contributions, and other backgrounds according to
different calculations. Also theory articles address (some of)
these issues. Since there is no single agreed standard, readers
are referred to the respective article for relevant details.

In this article we revive the old FK calculations, put in a modern 
context, and suggest that they can still be relevant. This is a 
continuation of the work begun in \cite{Sjostrand:2025qez}, but extends 
and corrects it, notably in the context of top width handling
in the generation chain. The issue here is that the FK Green's functions
integrate out the Breit--Wigner (BW) smearing of the two top masses, such
that the FK expressions are only functions of the $\t\tbar$ invariant mass.
Convenient to rapidly study cross sections as a function of this pair
mass, or equivalently as a function of energy above or below the threshold.
But in event generation also the event-by-event $\t$ and $\tbar$ masses
need to be selected, \ie sampling should be in three variables rather
than one only. As a solution to this, we will suggest that a more-or-less
complete unsmearing of the FK expressions allows a consistent combination
with BW-smeared top masses. 

The article is organized as follows. In Section~\ref{sec:formalism}
we review some basic notation, the Coulomb above-threshold enhancement
factors and the Green's functions, and show a few comparisons of basic
behaviour. Notably we clarify the origins of the Green's function 
expressions in the $\gammat \to 0$ limit, and how to address top mass 
smearing. Then we introduce a pragmatic and flexible approach for
the Monte Carlo generation of events according to these expressions,
implemented in the \Pythia \cite{Bierlich:2022pfr} event generator.  
In Section~\ref{sec:productionstudies} some results are presented,
which offer first hints how the model could fare in realistic comparisons
with data. Section~\ref{sec:decaystudies} introduces the implementation
of angular correlations in the decay chain of a pseudoscalar $\t\tbar$
state. Finally, Section~\ref{sec:summary} gives a summary and outlook.   

\section{The formalism}\label{sec:formalism}

The studies will be done for a nominal top mass of $\mt = 172.5$~GeV
and top width $\gammat = 1.34$~GeV, with top masses distributed
according to a (relativistic) Breit--Wigner \cite{Breit:1936zzb}, 
unless otherwise specified. The thus event-by-event-selected top
masses are called $\mtone$ and $\mttwo$. 

The standard leading-order (LO) matrix elements (MEs) (or, properly
speaking, $\d\hat{\sigma}/\d\that$) for $\g\g \to \t \tbar$ and 
$\q\qbar \to \t \tbar$ \cite{Combridge:1978kx} assume a common mass 
for $\t$ and $\tbar$, however, which in \Pythia event-by-event is
chosen to be 
\begin{equation}
\mtbar^2 = \frac{\mtone^2 + \mttwo^2}{2} -
\frac{ \left( \mtone^2 - \mttwo^2 \right)^2}{4\shat}
\end{equation}
such that
\begin{equation}
\betat = \sqrt{ \left( 1 - \frac{\mtone^2}{\shat}
  - \frac{\mttwo^2}{\shat} \right)^2 - 4 \, \frac{\mtone^2}{\shat}
  \, \frac{\mttwo^2}{\shat} }
  = \sqrt{1 - 4 \, \frac{\mtbar^2}{\shat}} ~. 
\label{eq:betat}
\end{equation}
Thus the three-momentum $\mathbf{p}_{\t 1} = - \mathbf{p}_{\t 2}$ 
in the $\t\tbar$ rest frame can be preserved with the temporary 
redefinition of $\that$ and $\uhat$
\begin{equation}
\overline{\that}, \overline{\uhat}
= - \frac{1}{2} \left\{ (\shat - 2\mtbar^2) 
\mp \shat \betat \cos\hat{\theta} \right\}
= (\that,\uhat) - \frac{ \left( \mtone^2 - \mttwo^2 \right)^2}{4\shat}~.
\end{equation}
Note that the phase space 
$\int \d \that = \int \d \overline{\that} \propto \betat$.
 
The LO MEs are combined with the NNPDF23\_lo\_as\_0130\_qed PDFs 
\cite{Ball:2013hta}, default in \Pythia since the Monash tune
\cite{Skands:2014pea}. Factorization and renormalization scales are 
both chosen as $Q^2 = \mt^2 + \pT^2$ with first-order 
$\as(m_{\Z}^2) = 0.13$ in MEs and PDFs alike.

The differential cross-section therefore schematically reads like
\begin{equation}
\d\sigma = \mathrm{BW}(\mtone) \, \d\mtone \, \mathrm{BW}(\mttwo) 
\, \d\mttwo \, f_i(x_1, Q^2) \, \d x_1 \, f_j(x_2, Q^2) 
\, \d x_2 \, \frac{\d\hat{\sigma}_{ij}}{\d\overline{\that}} \, 
\d\overline{\that} \, F_{\mathrm{mult}} ~,
\label{eq:mastergen}
\end{equation}
with $\shat = x_1 x_2 s$, and with $F_{\mathrm{mult}} = 1$ at Born level.

The $\q\qbar \to \t \tbar$ process involves an $s$-channel gluon
exchange, so is always in a colour octet state. For $\g\g \to \t \tbar$
both singlet and octet are possible, and we assume they occur in ratio
given by the colour factors
\begin{equation}
\frac{\mathrm{singlet}}{\mathrm{octet}}
= \frac{(\delta^{ab}/\sqrt{3})^2}{(d^{abc}/\sqrt{2})^2} = \frac{2}{5}~, 
\end{equation}
\ie a $2/7$ singlet fraction before the additional factors below are applied.

\subsection{The Coulomb factors}

Above threshold, the LO cross section is modified by multiple soft-gluon
exchanges. This is the QCD analogue of the same phenomenon in the Coulomb
potential of QED \cite{Sommerfeld:1931qaf,Gamow:1928zz,Sakharov:1948plh}.
For QCD the relevant resummed multiplicative factors $F_{\mathrm{mult}}$ in 
eq.~(\ref{eq:mastergen}) are
\begin{align}
|\Psi^{(s)}(0)|^2 &= \frac{X_{(s)}}{1 - \exp(-X_{(s)})}~~~\mathrm{with}~~~
X_{(s)} = \frac{4}{3} \frac{\pi\as}{\beta_{\t}}~, \\
|\Psi^{(8)}(0)|^2 &= \frac{X_{(8)}}{\exp(X_{(8)}) - 1}~~~\mathrm{with}~~~
X_{(8)} = \frac{1}{6} \frac{\pi\as}{\beta_{\t}}~,
\end{align}
where $(s)$ represent singlet and $(8)$ octet factors. The $\as$ scale
of gluon exchange is separate from that of $\t\tbar$ production, and 
is chosen to be 
\begin{align}
\as &= \as \left( \mt \sqrt{E^2 + \gammat^2} \right)~,~~~\mathrm{with}
\label{eq:alphas} \\
E &= \sqrt{\shat} - 2 \mt = \mhat - \mtone - \mttwo~.
\end{align}
Thus the scale is minimal at threshold, $E = 0$, where
$Q^2 = \mt \gammat \approx (15~\mathrm{GeV})^2$. In this case the
default is a second-order running $\as$ with $\as(m_{\Z}^2) = 0.118$,
unlike the LO ME choices, but can easily be changed.

\begin{figure}
\includegraphics[width=0.50\textwidth]{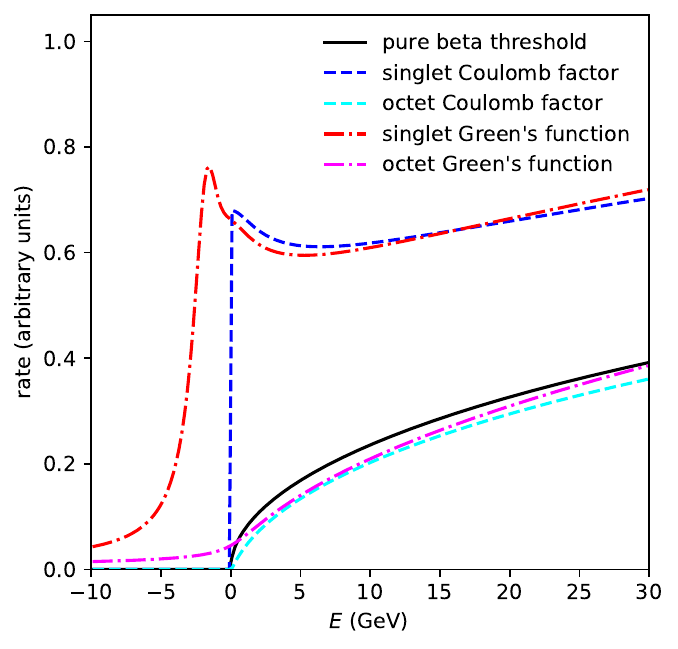}%
\includegraphics[width=0.50\textwidth]{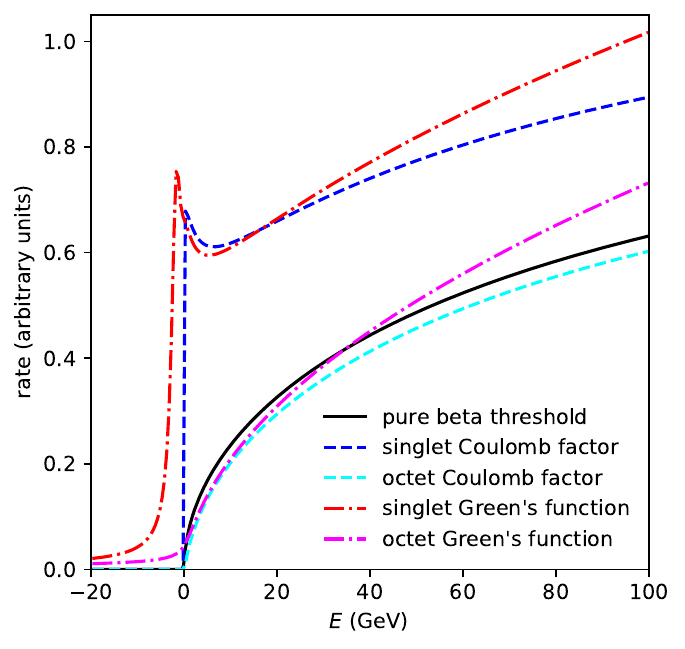}\\[-1mm]
\hspace*{0.25\textwidth}(a)\hspace{0.47\textwidth}(b)
\caption{Threshold $\betat$ behaviour of pure phase space, and
corrected by the Coulomb factors or replaced by the Green's functions, 
for a narrow or wide energy range, respectively.}
\label{fig:esimple}
\end{figure}

The resulting threshold behaviour is shown in Fig.~\ref{fig:esimple}.
The matrix elements are finite and slowly varying at threshold, so the 
phase space factor $\betat$ dominates the cross section variation. 
We then note that $|\Psi^{(s)}(0)|^2 \approx X_{(s)} \propto 1 / \betat$ 
such that $\betat |\Psi^{(s)}(0)|^2$ is finite at the threshold. 
By contrast $|\Psi^{(8)}(0)|^2 \to 0$ for $E \to 0$, thus suppressing
the threshold region cross section.

\subsection{The Green's functions}

The Green's function formalism allows to introduce the finite lifetime 
of the top quarks, using the NRQCD formalism of the top quarks 
propagating in a Coulomb potential. To obtain analytical expressions, 
a lowest-order potential $V(r) \propto \as / r$ is assumed in the FK
approach, with a 
constant $\as$. Once solutions have been found, however, the $\as$ in
these expressions are allowed to run, like in eq.~(\ref{eq:alphas}).
Further, the cross section is related to the imaginary 
part of the Green's functions $G_{E + i \gammat}$ at the origin,
and always occur with a prefactor $4\pi/\mt^2$, so it is convenient 
to define singlet $\tilde{G}^{(s)}(E)$ and octet $\tilde{G}^{(8)}(E)$ as  
\begin{align}
\tilde{G}^{(s)}(E) =
\frac{4\pi}{\mt^2} \Im G^{(s)}_{E + i\Gamma_{\t}}(0,0) &=
\frac{p_2}{\mt} + \frac{2p_s}{\mt} \arctan\frac{p_2}{p_1} \nonumber \\
&+ \frac{2 p_s^2}{\mt^2} \sum_{n=1}^{\infty} \frac{1}{n^4}
\frac{\gammat p_s n + p_2 \left(n^2 \sqrt{E^2 + \gammat^2}
  + \frac{p_s^2}{\mt} \right) }%
{\left( E + \frac{p_s^2}{\mt n^2} \right)^2 + \gammat^2} ~,
\label{eq:Gs} \\  
\tilde{G}^{(8)}(E) =
\frac{4\pi}{\mt^2} \Im G^{(8)}_{E + i\gammat}(0,0) &=
\frac{p_2}{\mt} + \frac{2p_8}{\mt} \arctan\frac{p_2}{p_1}
+ \frac{2 p_8^2}{\mt^2} \sum_{n=1}^{\infty}
\frac{\mt p_2}{(n p_1 - p_8)^2 + n^2 p_2^2} ~, \label{eq:G8}\\
\mathrm{where}~~ p_s &= \frac{2}{3} \mt \as ~,~~~\\
p_8 &= - \frac{1}{12} \mt \as ~,\\
p_{1,2} &= \sqrt{ \frac{\mt}{2} \left( \sqrt{E^2 + \gammat^2}
  \mp E \right) } ~.  
\end{align}
\begin{figure}
\hspace*{0.25\textwidth}%
\includegraphics[width=0.50\textwidth]{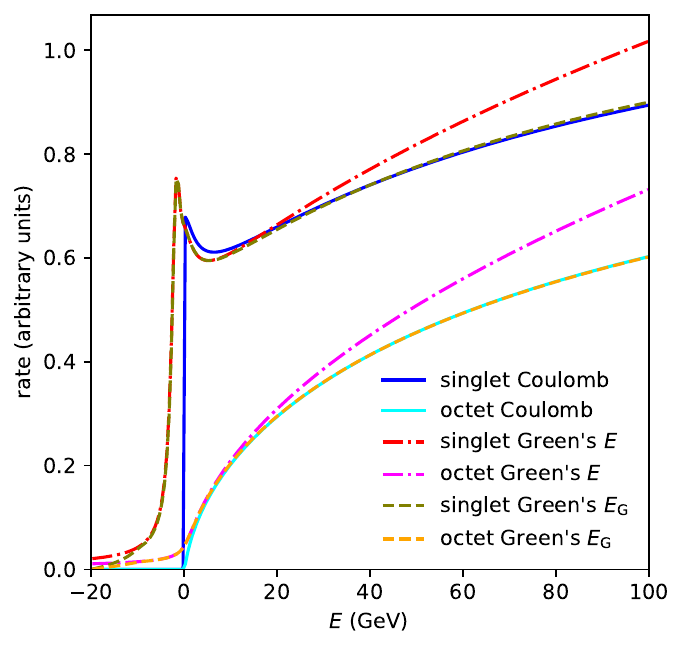}%
\caption{Green's functions with corrected argument $E_G$, compared
with initial argument $E$ and with Coulomb factor.}
\label{fig:EGargument}
\end{figure}

The $\tilde{G}(E)$ behaviour is shown in Fig.~\ref{fig:esimple}.
Note that there is a ``normalization'' difference between the
Coulomb and the Green's function formalisms, where the Coulomb
$|\Psi(0)|^2$ factors multiply the $\betat$ phase space,
while the $\tilde{G}$ expressions replace it, \ie
$F_{\mathrm{mult}} = \tilde{G} / \betat$ in eq.~(\ref{eq:mastergen}) , 
as we will return to later.

Notable in Fig.~\ref{fig:esimple}(b) is the divergence  of both
$\tilde{G}^{(s)}(E)$ and $\tilde{G}^{(8)}(E)$ for energies well 
above the threshold, where the Coulomb factors offer a more 
trustworthy extrapolation. The problem here is that $E$ is intended
to equate $\mt \betat^2$. That these agree for $E \approx 0$, where
$\mhat \approx 2\mt$, can be seen from 
\begin{align}
E &= \mhat - 2 \mt = (\mhat - 2 \mt) 
\left( \frac{\mhat +2\mt}{\mhat +2\mt} \right) 
= \frac{\shat - 4 \mt^2}{\mhat + 2 \mt}     \nonumber \\  
   &= \frac{\mhat^2}{\mhat + 2\mt} \left( 1 - 4\frac{\mt^2}{\shat} \right)
\approx m_{\t} \, \left(1 - 4\frac{\mt^2}{\shat} \right)
    = m_{\t} \, \betat^2 ~, \label{eq:Efrombetat}
\end{align}
but they gradually diverge for larger $E$. For two unequal masses the 
generalization of $\mt \betat^2$ is to introduce an alternative
\begin{equation}
E_G = \mtbar \, \left( \left( 1 - \frac{\mtone^2}{\shat}
  - \frac{\mttwo^2}{\shat} \right)^2 - 4 \, \frac{\mtone^2}{\shat}
  \, \frac{\mttwo^2}{\shat} \right) ~,
\end{equation}
and to use $E_G$ as argument for the Green's functions from here on,
rather than $E$.
(But figures are still shown as a function of $E$, not $E_G$.)
This then gives good agreement with the Coulomb behaviour for positive
$E$, almost perfectly so for $E > 10$~GeV, as can be seen in 
Fig.~\ref{fig:EGargument}. 

\subsection{Origins of the Green's functions}

It is interesting to understand the structure of the analytic
expressions for the Green's functions, and whether the numerical 
agreement with the the Coulomb expression for large $E$ is
coincidental. Unfortunately the original notes are no longer
available. What is known is that the top width was introduced into 
each quark propagator in the intermediate states in the gluon ladder, 
and then the integration over the final state was performed with
non-relativistic Breit--Wigners, using calculus of residues. 
This implies an integration over top masses between 
$-\infty$ and $+\infty$, so already there a difference relative to the 
\Pythia mass handling. It also implies a risk of double-counting
BW effects. This is unlike the Coulomb expressions, where no width
effects are included, so that any $\t$ and $\tbar$ mass smearing
entirely comes from the \Pythia mass selection.
 
A study of the limit $\gammat \to 0$ shows that the singlet Green's 
function is equal to the ditto Coulomb, modulo the $\betat$ prefactor, 
plus a series of bound states:
\begin{equation}
\tilde{G}^{(s)}(E) = \betat \,\frac{X_{(s)}}{1 - \exp(-X_{(s)})}
+ \frac{4 \pi p_s^3}{\mt^2} \sum_{n = 1}^{\infty} \frac{1}{n^3} \,
\delta \left( E + \frac{p_s^2}{\mt n^2} \right) ~.
\label{eq:greensdelta}
\end{equation}
Since there are no bound states in the octet channel, there the 
agreement between the Coulomb and the Green's function is perfect. 
Let us study the claimed equivalence in detail.

\begin{figure}
\includegraphics[width=0.50\textwidth]{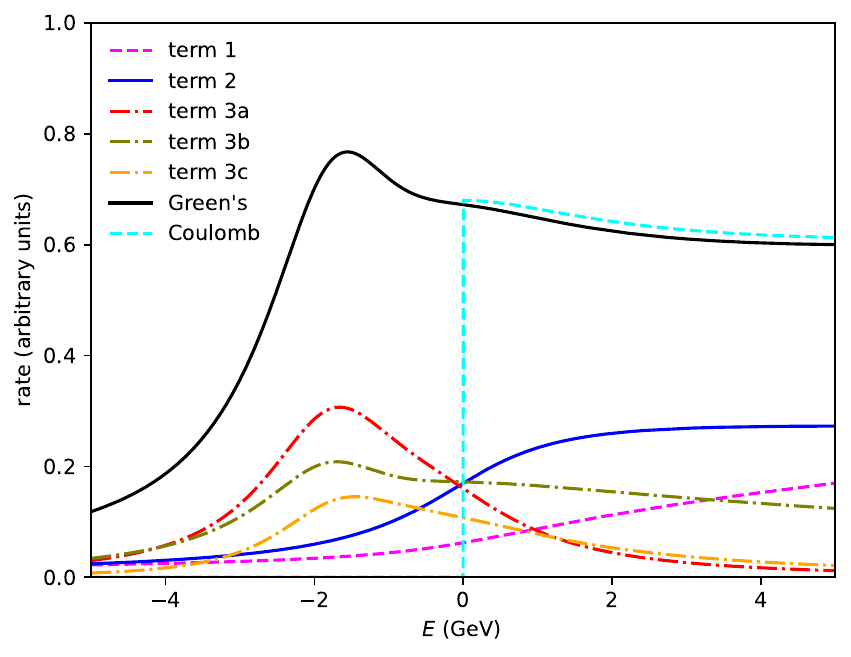}%
\includegraphics[width=0.50\textwidth]{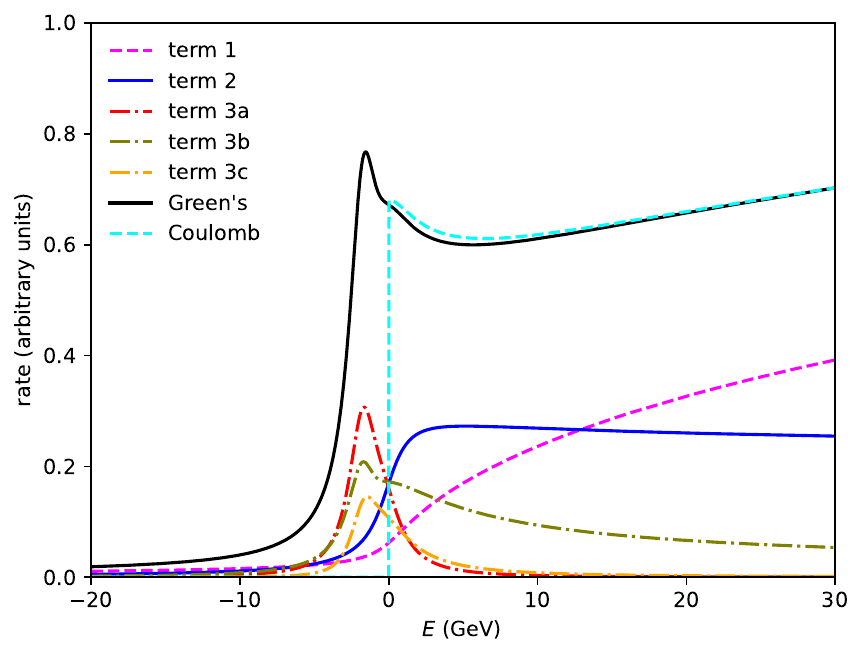}\\[-1mm]
\hspace*{0.25\textwidth}(a)\hspace{0.45\textwidth}(b)\\[2mm]
\includegraphics[width=0.50\textwidth]{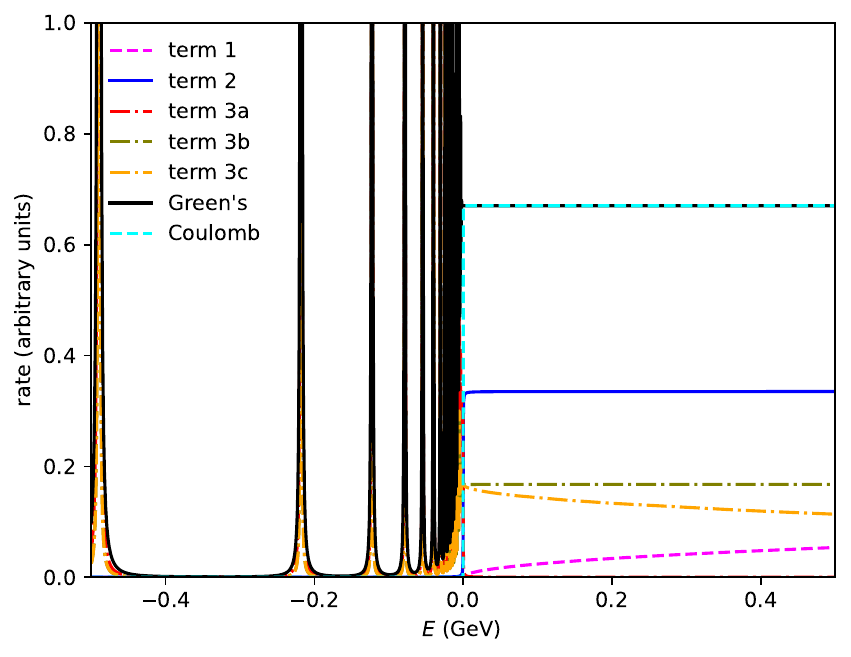}%
\includegraphics[width=0.50\textwidth]{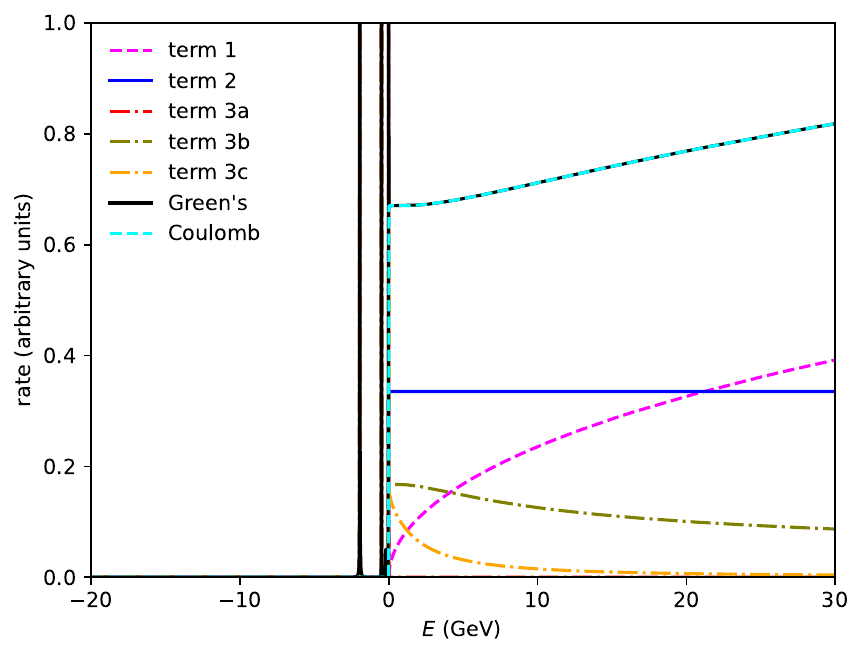}\\[-1mm]
\hspace*{0.25\textwidth}(c)\hspace{0.47\textwidth}(d)
\caption{Decomposition of the colour singlet Green's function,
as described in the text. Frames (a) and (b) for $\gammat = 1.34$~GeV, 
and (c) and (d) for $\gammat = 10^{-4}$~GeV. For the latter two frames 
the resonance peaks are truncated, the term 3 sub-terms visually overlap
for $E < 0$,  and the Green's and Coulomb curves overlap for $E > 0$.}
\label{fig:Gsindecomposed}
\end{figure}

The singlet case is the most interesting one, so we 
will concentrate on $\tilde{G}^{(s)}$, eq. (\ref{eq:Gs}). In this
expression, call $p_2 / \mt$ term 1, the $\arctan$ part term 2,
and the sum term 3. Within the latter, the part with a
$\gammat p_s n$ numerator is term 3a, the one with
$p_2 n^2 \sqrt{E^2 + \gammat^2}$ term 3b, and the one with
$p_2 p_s^2/\mt$ term 3c. A split into these terms is shown in 
Fig.~\ref{fig:Gsindecomposed}. Frames (a) and (b) are for the normal 
$\gammat = 1.34$~GeV, while (c) and (d) are for a scenario where
$\gammat = 10^{-4}$~GeV, as an approximation of the $\gammat \to 0$
limit. In the latter, a fixed $\as = 0.16$ is used, since else the 
ansatz in eq.~(\ref{eq:alphas}) would derail for $E \approx 0$.

One may then note that term 1 reduces to the $\betat$ threshold
factor of the Born cross section in the $\gammat \to 0$ limit,
and that it contains no $\as$ dependence. 
Term 2, with its $\arctan(p_2/p_1)$, reduces to a step function,
and is of $\mathcal{O}(\as)$. Term 3 is $\mathcal{O}(\as^2)$ and 
higher, and is the one that contains an infinite series of 
resonance states at $E_n = - p_s^2 /(\mt n^2)$.

At first glance it would seem that terms 3b and 3c do not contain
the expected $\gammat$ factor in the numerator. But, for $E < 0$, 
\begin{equation}
p_2 = \sqrt{ \frac{\mt}{2} \left( \sqrt{E^2 + \gammat^2} + E \right) } 
\approx \sqrt{ \frac{\mt}{2} \left( |E| + \frac{\gammat^2}{2 |E|} + E \right) }
= \gammat \sqrt{\frac{\mt}{4|E|}} ~,
\end{equation} 
so the whole numerator is proportional to $\gammat$. Now consider the 
first and most important term in the series, $n = 1$, for which the 
peak position is $E = - p_s^2 /\mt$. Then the numerator becomes
\begin{equation}
\gammat p_s + p_2 \left(\sqrt{E^2 + \gammat^2} + \frac{p_s^2}{\mt} \right)
\approx \gammat p_s + \frac{\gammat \mt}{2 p_s} 
\left( \frac{p_s^2}{\mt} + \frac{p_s^2}{\mt} \right) 
= \gammat p_s \left( 1 + \frac{1}{2} + \frac{1}{2} \right) ~.
\end{equation}
Noting that $|E_n| \propto 1/n^2$, all three sub-terms scale like
$n/n^4$, and thus keep their relative importance. Integrating the
Breit--Wigners gives the $\delta$ functions of eq.~(\ref{eq:greensdelta}).

For $E > 0$ the key is the ``mathematical trick'' \cite{Barbieri:1973lza}
\begin{equation}
\frac{z}{1 - e^{-z}} = 1 + \frac{z}{2}
+ 2 z^2 \sum_{n=1}^{\infty} \frac{1}{z^2 + (2\pi n)^2}~,
\label{eq:trick}
\end{equation}
This can be applied to the Coulomb expression, with
$z = X_{s} = (4/3) \pi \as / \betat = 2\pi \, p_s / \sqrt{m_t E}$,
where we used that $p_s = 2\mt\as / 3$ and $\betat = \sqrt{E / \mt}$,
to give
\begin{align}
|\Psi^{(s)}(0)|^2 = \frac{X_{(s)}}{1 - \exp(-X_{(s)})} 
&= 1 + \frac{X_{(s)}}{2} + 2 X_{(s)}^2
\sum_{n=1}^{\infty} \frac{1}{(2\pi n)^2 + X_{(s)}^2} \nonumber \\
&= 1 + \frac{2\pi\as}{3\betat} + 2 \frac{p_s^2}{\mt E}
\sum_{n=1}^{\infty} \frac{1}{n^2 + \frac{p_s^2}{\mt E}} \nonumber \\
&= 1 + \frac{2\pi\as}{3\betat} + 2 \frac{p_s^2}{\mt}
\sum_{n=1}^{\infty} \frac{1}{n^2} \, \frac{1}{E + \frac{p_s^2}{\mt n^2}} ~.
\end{align}
Now compare this with the Green's function for $\gammat = 0$,
noting that term 3a will not contribute, since it corresponds to
$\delta$ functions with $E < 0$. Both 3b and 3c will, however, since
$p_2$ switches from $\propto \gammat$ for negative $E$ to
$p_2 = \sqrt{\mt E} = \mt \betat$ (\cf eq.~(\ref{eq:Efrombetat}))
for positive $E$. Then
\begin{align}
\tilde{G}^{(s)}(E) &= \betat + \frac{2 p_s}{\mt} \, \frac{\pi}{2}
+ 2 \frac{p_s^2}{\mt^2} \sum_{n=1}^{\infty} \frac{1}{n^4} \,
\frac{p_2 \left( n^2 E + \frac{p_s^2}{\mt} \right)}%
{\left( E + \frac{p_s^2}{\mt n^2} \right)^2} \nonumber \\
& = \betat + \frac{2\pi\as}{3} + 2 \betat \frac{p_s^2}{\mt}
\sum_{n=1}^{\infty} \frac{1}{n^2} \, \frac{1}{E + \frac{p_s^2}{\mt n^2}}
= \betat \, |\Psi^{(s)}(0)|^2 ~.
\label{eq:Gsseries}
\end{align}

While the original Coulomb expression is easily evaluated, the 
infinite series expression is slowly converging for $E \to 0$
when a small $\gammat$ is used. In Fig.~\ref{fig:Gsindecomposed}(c,d) 
$10^4$ terms are used to get a convergence to the level of $10^{-4}$ 
down to $E = 0.1$~GeV. Convergence is faster for a realistic $\gammat$.  

At larger $E$, $E \gg p_s^2/\mt$, the final term of 
eq.~(\ref{eq:Gsseries}) simplifies to 
\begin{equation} 
2 \betat \frac{p_s^2}{\mt} \sum_{n=1}^{\infty} \frac{1}{n^2} \, \frac{1}{E}
= 2 \betat \frac{(2\mt\as/3)^2}{\mt^2\betat^2} \, \frac{\pi^2}{6}
= \frac{4}{27} \, \frac{\pi^2 \as^2}{\betat}~,
\end{equation}
so that
\begin{equation}
\tilde{G}^{(s)}(E) \approx \betat + \frac{2\pi\as}{3} + 
\frac{4\pi^2\as^2}{27\betat}
\end{equation}
for large $E$. This can be compared with the Coulomb power-series 
expansion
\begin{align}
\betat \, |\Psi^{(s)}(0)|^2 &= \betat \, \frac{X_{(s)}}{1 - \exp(-X_{(s)})}
= \betat \left( 1 + \frac{1}{2} X_{(s)} + \frac{1}{12} X_{(s)}^2
- \frac{1}{720} X_{(s)}^4  + \cdots \right) \nonumber \\
&= \betat + \frac{2\pi\as}{3} + \frac{4\pi^2\as^2}{27\betat} + \cdots
= \tilde{G}^{(s)}(E) + \cdots
\end{align}
where $X_{(s)}$ is below unity for not too small $\betat$, so the
truncation after the third term is justified. (A similar observation 
was made in \cite{Fadin:1995fp} for the case of purely QED final-state 
interaction caused by the photon ladder diagrams in the near-threshold 
$\W^+\W^-$ pair production in the $\e^+\e^- \to \W^+\W^-$ process.)
Note that $X_{(s)}$ blows up for small $E$, so there this expansion 
approach is badly divergent, with varying signs for the terms. 
By comparison eq.~(\ref{eq:trick}) consists only of positive terms, 
so convergence is under control. 

It may seem like a coincidence that eq.~(\ref{eq:trick}) leads to 
denominators of the Green's function series that directly relate to the 
$E_n$ bound-state energies, but more likely it reflects the analytic 
properties of the Green's function in the zero-width limit
in the complex energy plane. Whatever the origin, it allows the 
below-threshold bound states and above-threshold Coulomb expansion 
to be combined into a series with a common denominator structure, 
which are even closer intermixed after the analytical convolution with 
Breit--Wigners, thereby masking the origin of the full expression.  

\begin{figure}
\includegraphics[width=0.50\textwidth]{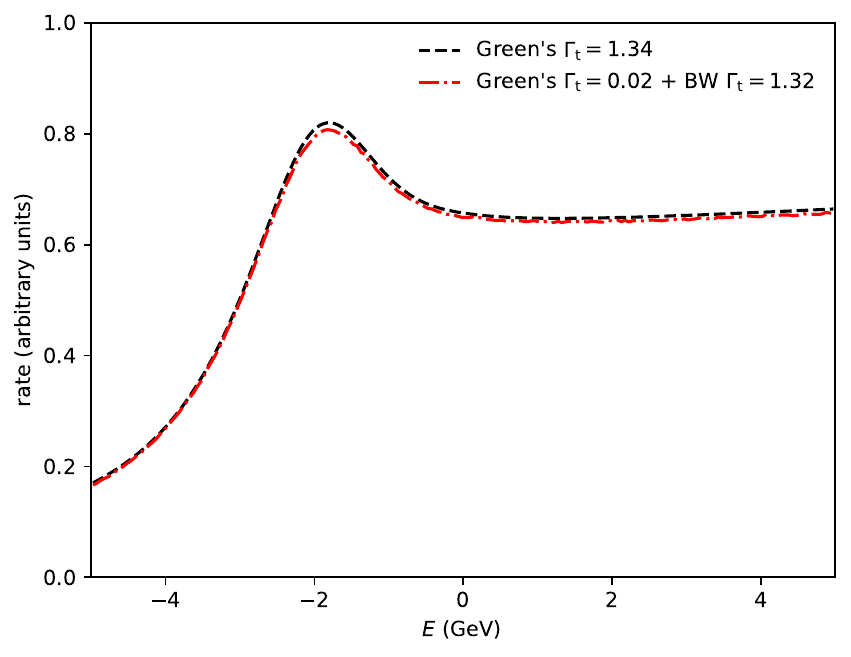}%
\includegraphics[width=0.50\textwidth]{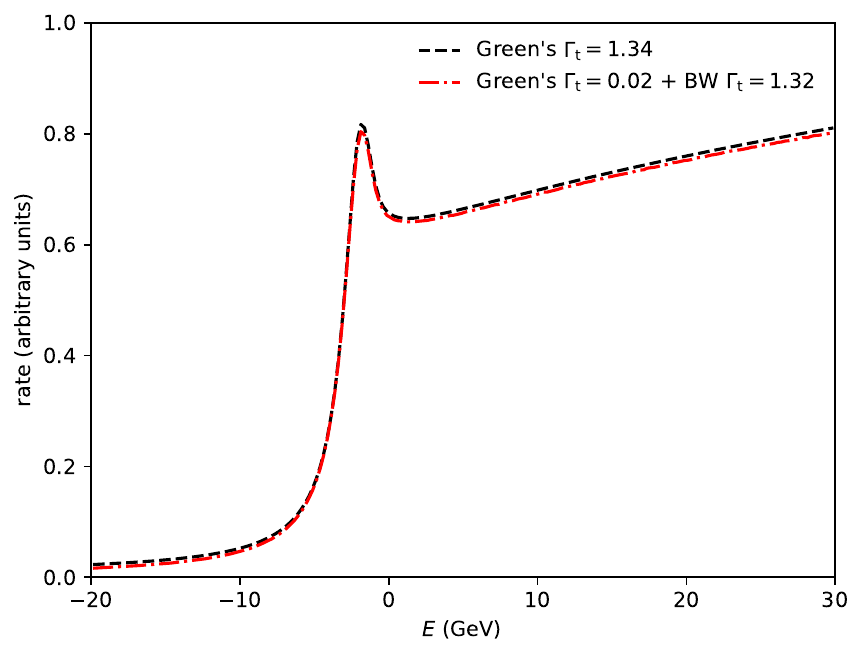}\\[-1mm]
\hspace*{0.25\textwidth}(a)\hspace{0.45\textwidth}(b)
\caption{Comparison of the standard Green's function expression using
$\gammat = 1.34$~GeV with one where $\gammat = 0.02$~GeV but this then
is smeared with $\t$ and $\tbar$ non-relativistic Breit-Wigners having 
$\gammat = 1.32$~GeV. A fix $\as = 0.16$ is used.}
\label{fig:GsinSmear}
\end{figure}

A check of our conclusions is to smear the Green's function in the 
$\gammat \to 0$ limit with Breit--Wigners for the $\t$ and $\tbar$ 
masses. Non-relativistic BWs obey additivity of widths, 
\ie the convolution of two BWs with widths $\Gamma_1$ and
$\Gamma_2$ gives a new BW with width $\Gamma_1 + \Gamma_2$. Therefore 
in Fig.~\ref{fig:GsinSmear} we compare the standard Green's function
with a narrow one, where instead Monte-Carlo-generated BW $\t$ and
$\tbar$ masses are used to add up to the same total width. Overall
the agreement is good, and we assume the remaining minor discrepancies
can be attributed \eg to how the BW tails are handled.
  
\begin{figure}
\includegraphics[width=0.50\textwidth]{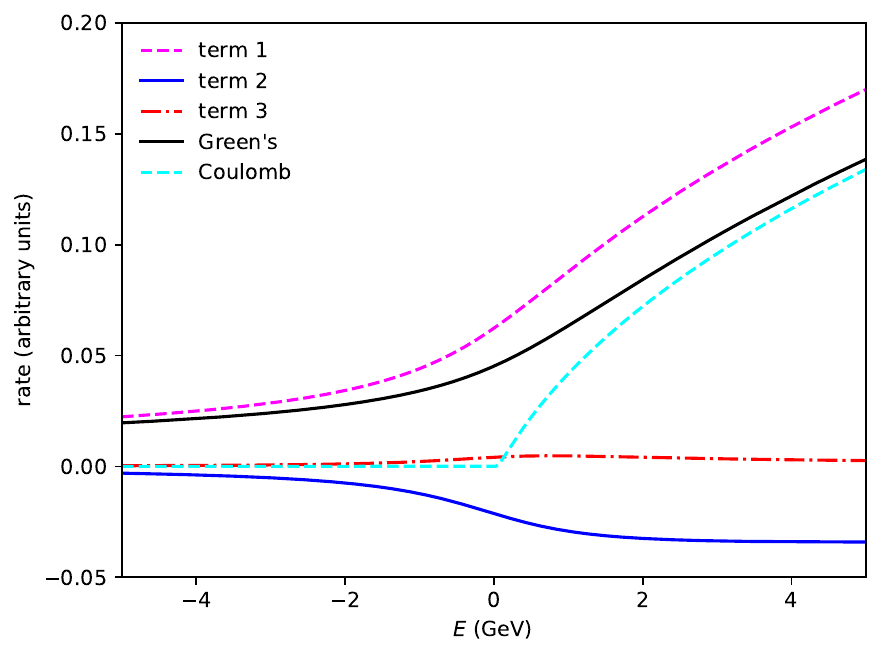}%
\includegraphics[width=0.50\textwidth]{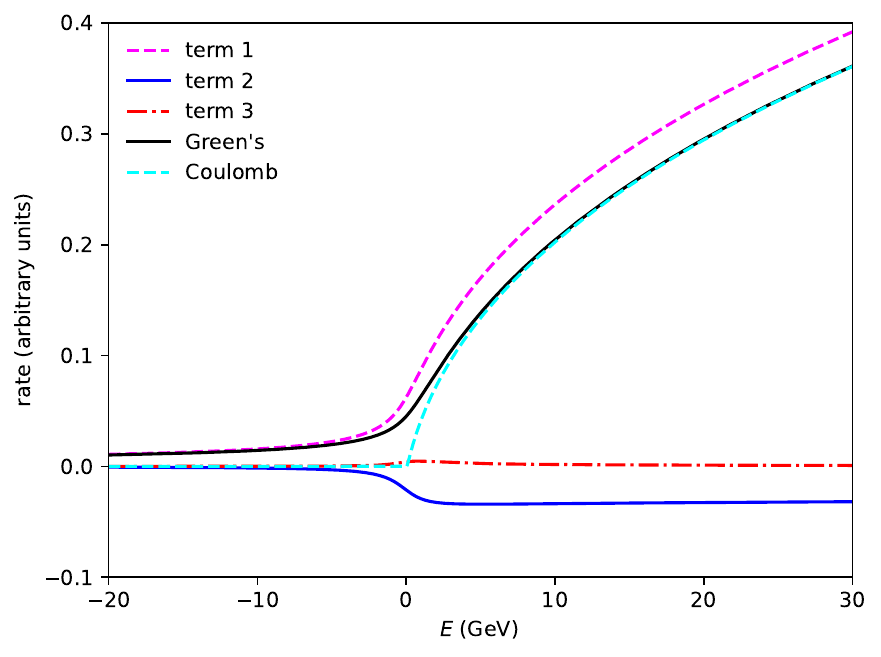}\\[-1mm]
\hspace*{0.25\textwidth}(a)\hspace{0.45\textwidth}(b)\\[2mm]
\includegraphics[width=0.50\textwidth]{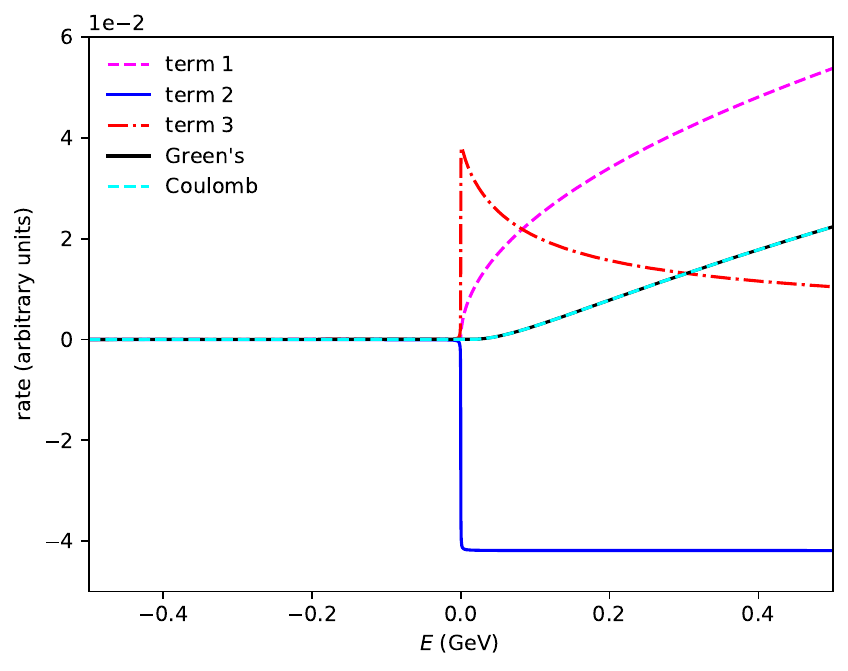}%
\includegraphics[width=0.50\textwidth]{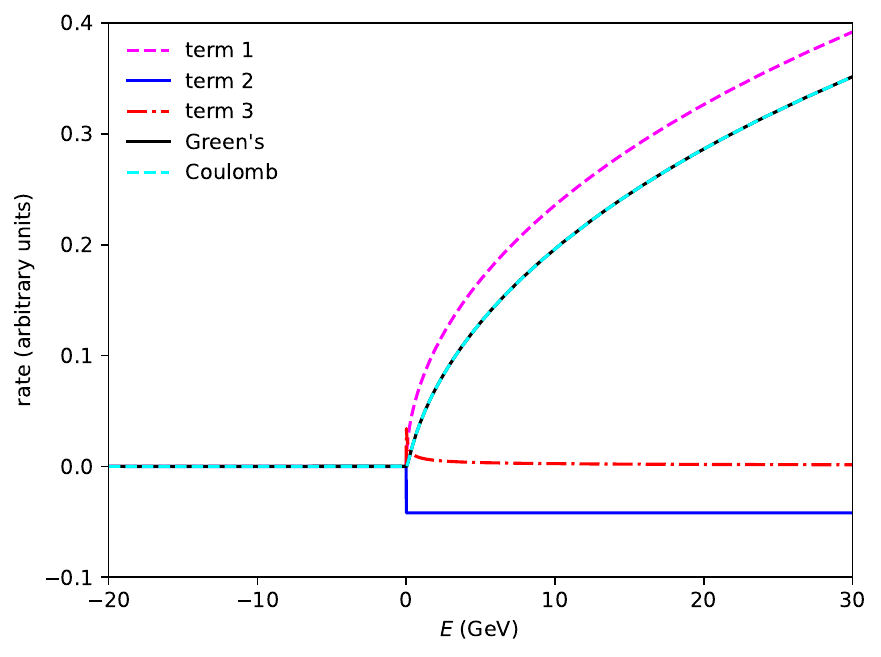}\\[-1mm]
\hspace*{0.25\textwidth}(c)\hspace{0.47\textwidth}(d)
\caption{Decomposition of the colour octet Green's function,
as described in the text. Frames (a) and (b) for $\gammat = 1.34$~GeV, 
and (c) and (d) for $\gammat = 10^{-4}$~GeV. The Green's and Coulomb
curves overlap for the latter two frames.}
\label{fig:Goctdecomposed}
\end{figure}

For the colour octet case the story repeats, Fig.~\ref{fig:Goctdecomposed}. 
Considering the Green's function expression,
eq.~(\ref{eq:G8}), note that $p_1 = 0$ for $\gammat = 0$ and $E > 0$, 
and that $p_2^2 = m_t E$, such that the denominators can be rewritten 
as $E + p_8^2 / (\mt n^2)$. The analogy with the singlet case then is 
apparent, and again the use of eq.~(\ref{eq:trick}) may trick you into 
assuming that the denominators are related to the presence of 
below-threshold bound states, which is not the case. It is also 
interesting to note that this ``term 3'' series gives a tiny 
contribution, visible just above threshold for a small $\gammat$, 
but negligible when smeared by realistic $\gammat$. Relative to the 
singlet case, this is because $|p_8| / p_s = X_{(8)} / X_{(s)} = 1/8$, 
which enters quadratically for term 3.  

\subsection{Event generation}

Generation of events above threshold, $E > 0$, is fairly 
straightforward. First $\mtone$ and $\mttwo$ are selected
according to Breit--Wigners, and based 
on that $\mtbar$ and $\mhat_{\mathrm{min}} = \mtone + \mttwo$ are
defined. This allows a selection of $(x_1, x_2, \that)$, biased to 
improve efficiency, within the allowed phase space, and an
accept/reject step based on PDFs, (LO) MEs and the 
$F_{\mathrm{mult}}$ factor, $|\Psi^{(s,8)}(0)|^2$ for the Coulomb 
case and $\tilde{G}^{(s,8)}(E_{G}) / \betat$ for the Green's function
one. For a $\g\g$ initial state the weight is $2/7$ the singlet one
and $5/7$ the octet one, while a $\q\qbar$ initial state is pure octet.

The extension to $E < 0$ is more tricky. A first issue is that
the Green's function expressions have tails that extend to arbitrarily small
$E$. Since the formalism has been derived for $E$ close to 0, 
such tails could not be trusted. Therefore we have chosen to 
apply Green's function weights unmodified only down to $E = -10$~GeV, and 
thereafter  damp them linearly down to $E = -20$~GeV, so that they 
vanish below that.

This means that, for a given preliminary 
$\mhat_{\mathrm{min}} = \mtone + \mttwo$, the phase space initially is 
populated down to 
$\mhat_{\mathrm{min}}' = \mhat_{\mathrm{min}} - 20$~GeV, notably
with an $\mhat > \mhat_{\mathrm{min}}'$.
When $E = \mhat - \mtone - \mttwo > 0$ the normal procedure above 
is applied. When instead $E < 0$, two new $\mtone' < \mtone$ and 
$\mttwo' < \mttwo$ are selected, repeatedly until 
$E' = \mhat - \mtone' - \mttwo' > 0$. Based on these masses, 
a new $\betat'$ is calculated, and the event is assigned a ``hybrid''
weight $\tilde{G}^{(s,8)}(E_{G}) / \betat'$. Here the numerator 
Green's functions are damped for $E < 0$ as already described, 
while the denominator cancels against the $\betat'$ factor in the 
$\that$ part of the phase space defined by $\mhat$, $\mtone'$ and 
$\mttwo'$.

This solution is intended to push the $\t$ and $\tbar$ off-shell,
in a way that represents the BW factors, times 
ME, PDF and phase-space effects. It is relevant here to note that, 
whether above threshold or in a pseudo-bound ``toponium'' state, 
a perfect theorist's detector would record the $\b$ and $\W$ decay 
products and from that reconstruct $\t$ and $\tbar$ masses. In that
sense the separation of ``ordinary'' BW mass smearing and 
below-threshold production is ill-defined on an event-by-event basis, 
but is reflected in distorted inclusive top mass spectra, singly 
and for the pair.
 
Finally, an important update. As we have seen, the FK Green's function
expressions involve an inclusive smearing of top masses by smart
tricks and calculus of residues. This is convenient for numerical
studies where the individual top masses are not the issue, neither
above nor below threshold. But it clashes with the generation of a
complete event, where those top masses should be selected   
event-by-event. At the time of first reviving the old 
calculations \cite{Sjostrand:2025qez}, the inclusion of the top width 
in the Green's functions had not been sorted out, however. Therefore 
\Pythia first generated Breit--Wigner-smeared $\t$ and 
$\tbar$ masses, and then applied $\tilde{G}^{(s,8)}$ weights that 
themselves had already been convoluted with BW factors, 
leading to a double-counting of mass smearing. Some possible solutions 
to avoid this are as follows.
\begin{enumerate}
\item Include mass smearing for the $\mtone$ and $\mttwo$ selection, 
but combine that with Green's function with $\gammat = 0$, 
eq.~(\ref{eq:greensdelta}) and its octet equivalent. The disadvantage 
here is mainly technical, that it becomes necessary to combine 
a discrete phase space (with infinitely many peaks) below threshold
with a continuous one above it. \Pythia is not prepared for such an
option, so that would require nontrivial extra work. 
\item Fix $\mtone = \mttwo = \mt$ and use the mass-smeared Green's 
functions, eqs.~(\ref{eq:Gs}) and (\ref{eq:G8}), as event weight. 
This would give the desired pair invariant mass distribution above 
threshold, but not the smeared individual top masses. Approaching
threshold $\betat$ would vanish. Paradoxically, this approach
would still require that $\mtone'$ and $\mttwo'$ are selected with a 
width, in order to sample the $E < 0$ phase space, so the
below-threshold description would be more realistic than the
above-threshold one. In short, this approach is convenient to plot
expected $E$ spectra when the individual top masses are not studied,
as shown in the figures so far, but not to generate complete events.
\item We already noted the additivity of widths in the convolution of 
Breit--Wigners. This means it is possible to interpolate
between the two extremes above, \ie generate $\mtone$ and $\mttwo$ 
with a width $\gammatBW$, and then weigh with Green's functions 
where $\gammatG = \gammat - \gammatBW$. As before, $\mtone'$ and 
$\mttwo'$ are selected with the full $\gammat$ width, and also 
$\as$ in eq.~(\ref{eq:alphas}) uses this. For a small $\gammatG$
this is almost equivalent with option 1, but avoids the technical
complications.
\end{enumerate}

The main user switch in the code, to address these issues, is
\texttt{TopThreshold:model}. Option 0 means pure Born-level $\t\tbar$
production, and 1 includes the Coulomb factors, in both cases with
normal top mass smearing without any double-counting issues. Then 2
allows the top quark width and the Green's function width to be set 
separately, whereas the preferred option 3 reduces the generated top
quark width by the width assigned to the Green's function. The latter
then gives the intended total smearing of the threshold behaviour.
See the \texttt{html} manual page \texttt{Top}, section 
\texttt{Threshold enhancements} for a description of further 
settable parameters.

We end this section with a note on the evolution of the \Pythia
code. The Fortran code, available from 1990 onwards
\cite{Sjostrand:1993yb}, only implemented the Coulomb factors
for above-threshold enhancement, while below-threshold cross
sections were obtained by numerical integration rather than
Monte Carlo generation. This Coulomb implementation was not
carried over to the \Pythia~8 C++ code. Instead it was first
re-implemented in \Pythia~8.316 from October 2025, along with first
implementations of Green's functions, valid also below
threshold, but suffering from double-counting of the top width
as already discussed. In \Pythia~8.317 from January 2026 it
became possible to set widths separately for the Green's functions
and the top quark mass selection, to remove double-counting.
Angular correlations in the decay of a $\t\tbar$ pseudoscalar
state were also introduced, see below Sec.~\ref{sec:decaystudies}.
In the upcoming \Pythia~8.318 the new \texttt{TopThreshold:model = 3}
should make it easier to set up the two top mass usages consistently.
In addition some further information and options of a technical
nature are made available.

\section{QCD production studies}\label{sec:productionstudies}

Among the options discussed above, number 3 with a small $\gammatG$ 
is especially interesting, since it offers a rather clean separation 
of what is below- and above-threshold production. The limitation is 
that a small $\gammatG$ gives narrow and high $E < 0$ peaks, which 
are not caught in the \Pythia initialization, and therefore later give 
events with weights above unity, reducing the statistical power of an 
event sample. A $\gammatG \geq 0.1$~GeV only gives a few events with 
weights only modestly above unity. Even if the weights by mistake were 
omitted, the cross section in the resonance region would be 
underestimated by about 1\%, with smaller effects elsewhere.
Thus this would be a recommended intermediate solution. To address 
the threshold separation issue, however, we will here go as low as
$\gammatG = 0.02$~GeV, where the average weights contribute up to a
factor 1.4 in the threshold region, with a tail of weights stretching
up to 10. This is compensated by using higher statistics than
otherwise would have been needed; unless otherwise indicated 
$4 \cdot 10^8$ events are generated per scenario.

\begin{table}[tp!] \centering
\begin{tabular}{|l|cccc|}
\hline
& \multicolumn{4}{c|}{$\sigma$ (pb)} \\ \hline
model           & $E < 0$ & $\mhat < 345$ & $\mhat < 380$ & all $\mhat$ \\
\hline
Born            &  0.00   & 2.23 &  61.32   & 516.26 \\
Coulomb         &  0.00   & 3.21 &  74.35   & 558.57 \\
\textbf{narrow Green}  &  \textbf{4.41}   & \textbf{6.50}
&  \textbf{78.57}   & \textbf{562.78} \\
narrow top      &  6.66   & 6.69 &  80.75   & 570.08 \\ 
both wide       &  6.51   & 8.60 &  80.69   & 564.71 \\ \hline
$\g\g$ only     &  4.41   & 6.18 &  68.53   & 500.33 \\
$\q\qbar$ only  &  0.01   & 0.32 &  10.04   & 62.45 \\ \hline
$\g\g$ singlet + $\q\qbar$ octet
                & 15.36   & 18.17 & 139.05 & 744.27 \\
$\g\g$ octet + $\q\qbar$ octet
                & 0.04    &  1.84 &  54.39 & 490.18 \\ \hline
ME $\as^{(1)} = 0.130 \to \as^{(2)} = 0.118$
                & 3.67 & 5.41 & 65.38 & 469.46 \\
PDF NNPDF 2.3 LO $\to$ PDF4LHC21 NNLO
                & 4.47 & 6.62 & 81.14 & 581.72 \\ \hline
\end{tabular}
\caption{Cross section, partial and total, in various scenarios.
The first five are separate settings, while the last six are parts
or variations of the ``narrow Green'' scenario.}
\label{tab:sigmas}
\end{table}

In the studies of this section, the main scenario will be a
``narrow Green'' one, where $\gammatG = 0.02$~GeV and thus 
$\gammatBW = 1.32$~GeV for a total $\gammat = 1.34$~GeV. 
An alternative ``narrow top'' flips the balance to 
$\gammatG = 1.32$~GeV and $\gammatBW = 0.02$~GeV. 
As a reference to earlier studies \cite{Sjostrand:2025qez}, 
the ``both wide'' sets $\gammatG = \gammatBW = 1.34$~GeV.
In addition the Born and Coulomb scenarios offer simpler alternatives. 
But we have seen that the latter actually is closely related to
the above-threshold part of the Green's function, so offers a 
useful reference for the below-threshold resonance formation.

Cross-sections for these five options are shown in Table~\ref{tab:sigmas}.
The ``theoretical'' below-threshold contribution,
$E = \mhat - \mtone - \mttwo < 0$, obviously vanishes for the
Born and Coulomb options. But, by the Breit--Wigner smearing, they
still have non-vanishing cross-sections for the ``experimental''
$\mhat < 2\mt = 345$~GeV below-threshold definition. One may also
notice that the Coulomb correction increases the cross-section not
only around the threshold but also further above it.  

The narrow-Green 4.41~pb number for $E < 0$ represents the best
prediction of the true pseudo-bound-state contribution. This excess
relative to the Coulomb number is propagated to the total cross
section (``all $\mhat$''), within statistical errors. Instead the 
6.50~pb cross section for $\mhat < 345$~GeV  comes closer to an observable 
bound-state effect. Note that some events with $E < 0$ may still have
started out with a large $\mtone + \mttwo$, and thus $\mhat > 345$~GeV.  
That is, events flow in both directions between the $E < 0$ and 
$\mhat < 345$~GeV classifications.

In the option with a narrow top instead $E < 0$ and $\mhat < 345$~GeV
numbers agree more closely, owing to the decreased migration rate
from top mass smearing, and the $\mhat < 345$~GeV cross section agrees
reasonably well with the narrow-Green one, as should be expected.
There is a small cross-section increase, however, which is also present
for larger $\mhat$. Not unexpectedly the both-wide option gives the
largest $\mhat < 345$~GeV rate, since both the Green's function expression 
and Breit--Wigners here contribute to low-$\mhat$ production. But, as we
have discussed, this involves double-counting. 
 
In addition, Table~\ref{tab:sigmas} contains six variations 
of the main narrow-Green one. The first two separate the cross 
section into the contributions from the $\g\g \to \t\tbar$ and 
$\q\qbar \to \t\tbar$ channels, respectively. Overall the $\g\g$ 
one is an order of magnitude above the $\q\qbar$ one, and even larger
at small $E$ and $\mhat$, where only  $\g\g$ enjoys the colour-singlet
enhancement factor. In the next two, the $\g\g$ channel is varied
between pure singlet and pure octet (while $\q\qbar$ remains octet),
illustrating the sensitivity to the default assumed $2/7$ singlet
fraction. The last two illustrate the uncertainty already from the
simulation of the $\g\g \to \t\tbar$ and $\q\qbar \to \t\tbar$
Born-level cross sections. The default NNPDF 2.3 QCD+QED LO PDF
\cite{Ball:2013hta} comes with a first-order $\as^{(1)}(m_{\Z}) = 0.130$.
If changed to a second-order $\as^{(2)}(m_{\Z}) = 0.118$, not unexpectedly
cross sections overall drop by $\sim$17\%. (The Coulomb and Green's
function factor usage of an $\as^{(2)}(m_{\Z}) = 0.118$ with the scale 
choice of eq.~(\ref{eq:alphas}) remains unchanged, but obviously
offers yet another possible variation.) Finally, the PDF is replaced 
by the PDF4LHC21 NNLO one \cite{PDF4LHCWorkingGroup:2022cjn},
but still with $\as^{(1)}(m_{\Z}) = 0.130$.
Results are surprisingly similar between the older LO and newer NNLO
PDF. This may partly be a coincidence, but also partly that the
probed $x$ range of threshold-region $\t\tbar$ production, typically
$x \sim 0.01 - 0.1$, is in a reasonably well understood region,
more so than the $x \to 0$ and $x \to 1$ limits.

The recent article by Garzelli \etal \cite{Garzelli:2026ctb}
advocates the cross section in the 340~GeV $< \mhat <$ 350~GeV
as a convenient measure for comparisons. Their study of various
uncertainties lands at a value $11.67^{+1.43}_{-1.47}$~pb which,
after subtraction of non-resonant background cross section,
gives an excess of $4.15^{+1.43}_{-1.47}$~pb. To compare, our
preferred scenario above gives 11.00~pb, which reduces to
3.49~pb after subtraction of the Coulomb ``background'' scenario.
Results thus are compatible within errors.

\begin{figure}
\includegraphics[width=0.50\textwidth]{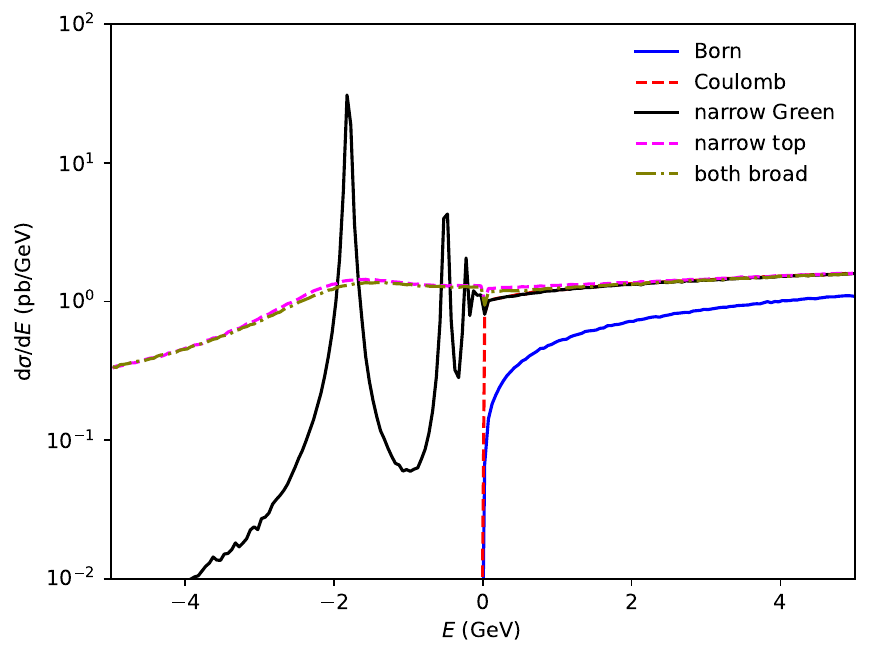}%
\includegraphics[width=0.50\textwidth]{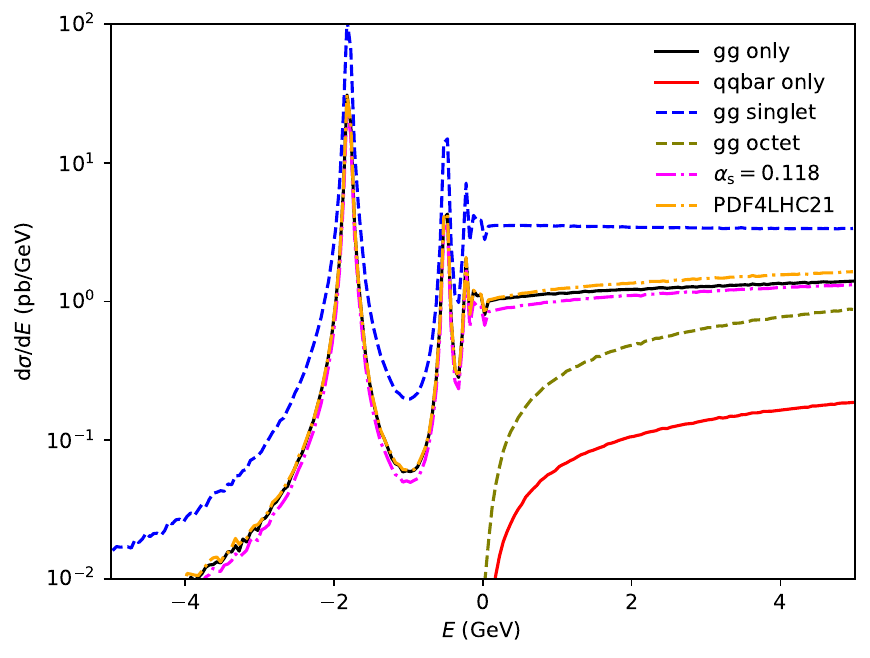}\\[-1mm]
\hspace*{0.25\textwidth}(a)\hspace{0.45\textwidth}(b)\\[2mm]
\includegraphics[width=0.50\textwidth]{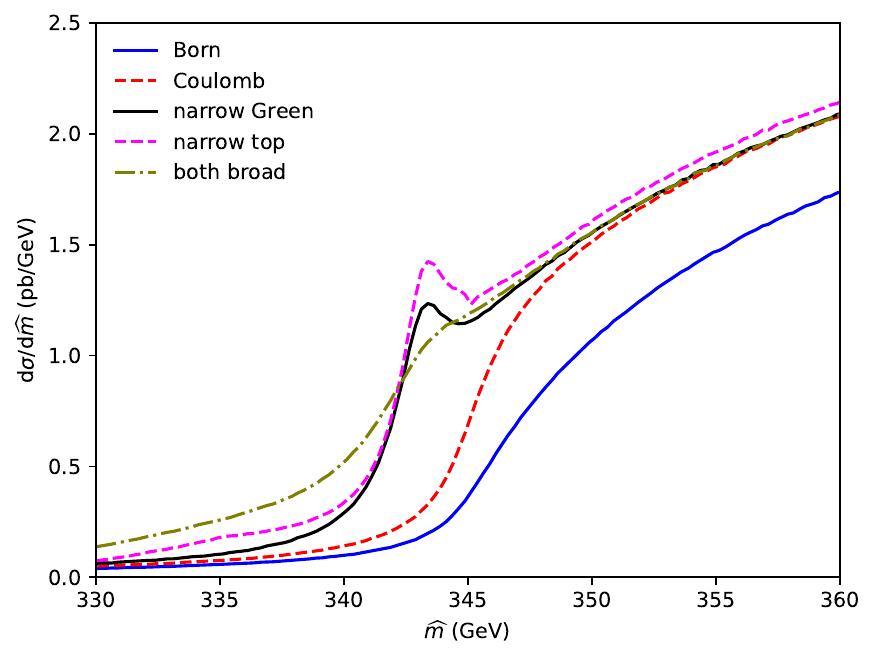}%
\includegraphics[width=0.50\textwidth]{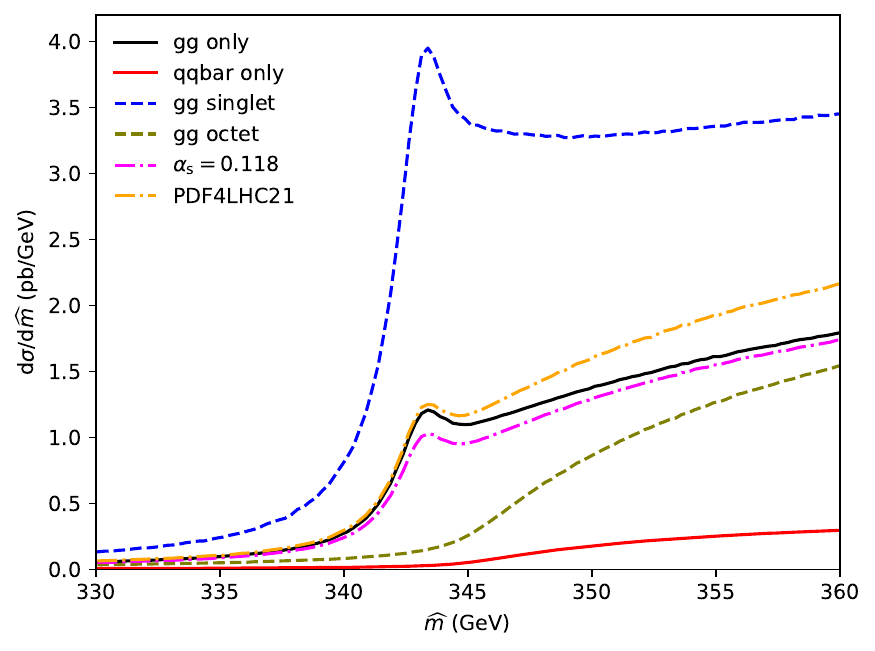}\\[-1mm]
\hspace*{0.25\textwidth}(c)\hspace{0.47\textwidth}(d)\\[2mm]
\includegraphics[width=0.50\textwidth]{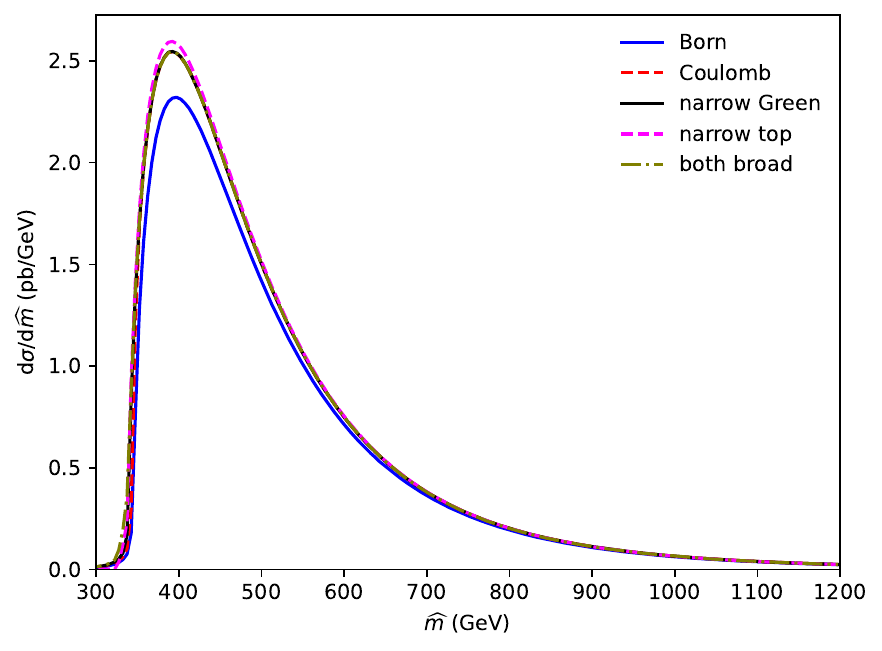}%
\includegraphics[width=0.50\textwidth]{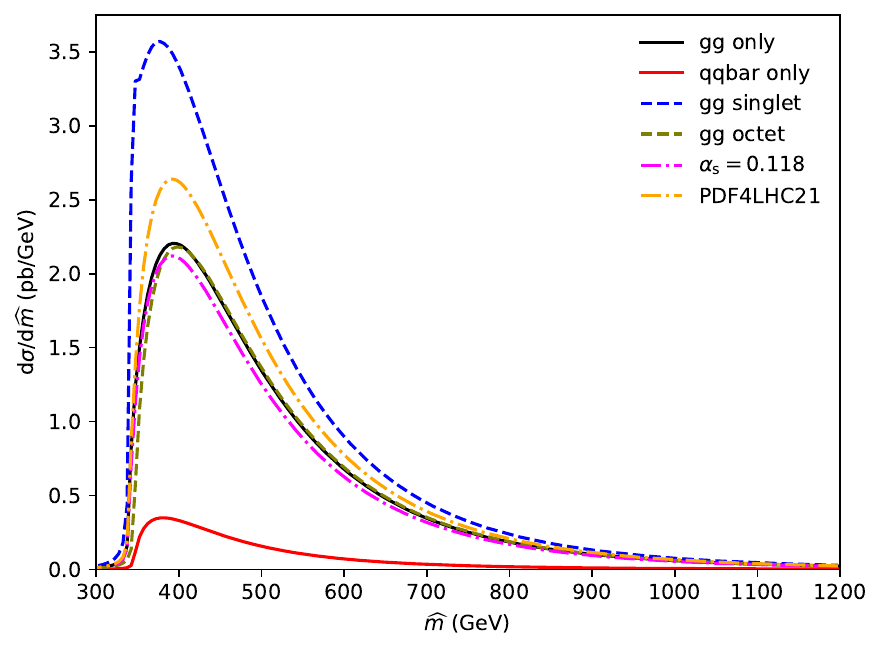}\\[-1mm]
\hspace*{0.25\textwidth}(e)\hspace{0.47\textwidth}(f)
\caption{Production cross section as a function of (a,b) threshold
energy $E$, (c,d) $\t\tbar$ invariant mass $\mhat$ in the threshold
region, and (e,f)  $\t\tbar$ invariant mass $\mhat$ over a broader
energy range.}
\label{fig:eandm}
\end{figure}

Fig.~\ref{fig:eandm} show some distributions of the 11 scenarios
described above, providing a differential view of the cross section
numbers in Table~\ref{tab:sigmas}. Frame (a) illustrates the
sharp resonance peaks obtained in the narrow-Green scenario, where
$\gammatG = 0.02$~GeV, necessitating a logarithmic $y$ scale.
In the narrow-top and both-broad scenarios the broader Green's
function smears out the individual peaks.

In frame (c) both the narrow-Green and narrow-top scenarios show
a small below-threshold peak, somewhat more pronounced 
for the latter. Visually, it would seem that the two differ in
$\mhat < 345$~GeV area somewhat more than implied by
Table~\ref{tab:sigmas}. One should here recall that the Green's
function expressions are killed 20~GeV below threshold, since the
expressions are only trustworthy in the threshold region. In the 
modelling the top Breit--Wigner is assumed relevant almost all the
way down to the $\W + \b$ threshold, on the other hand, giving a longer
low-$\mhat$ tail for the narrow-Green option. At the end of the day
the experimental cuts, imposed \eg to avoid contamination from
non-resonant $\W^+\b\W^-\bbar$ channels, will determine how much
these two options differ. The double-counted smearing in both-broad
removes any signs of a resonance enhancement, and also gives a higher
tail for negative $E$.

Frame (e) takes a broader view of the top cross section as a function
of $\mhat$. In the peak region the Born cross section is the lowest,
with the other four rather close, mainly being driven by the same
Coulomb enhancement. At larger $\mhat$ this enhancement dampens out,
and all five curves converge.

Fig.~\ref{fig:eandm} frames (b), (d) and (f) repeat the same study
for the six variants of the narrow-Green main scenario. The two curves
that only contain octet production die out at small $E$ and $\mhat$,
although an occasional $E < 0$ event is permitted by the 0.02~GeV
width in the Green's function. The other curves largely are rescaled
versions of the baseline scenario.
  
\begin{figure}
\includegraphics[width=0.50\textwidth]{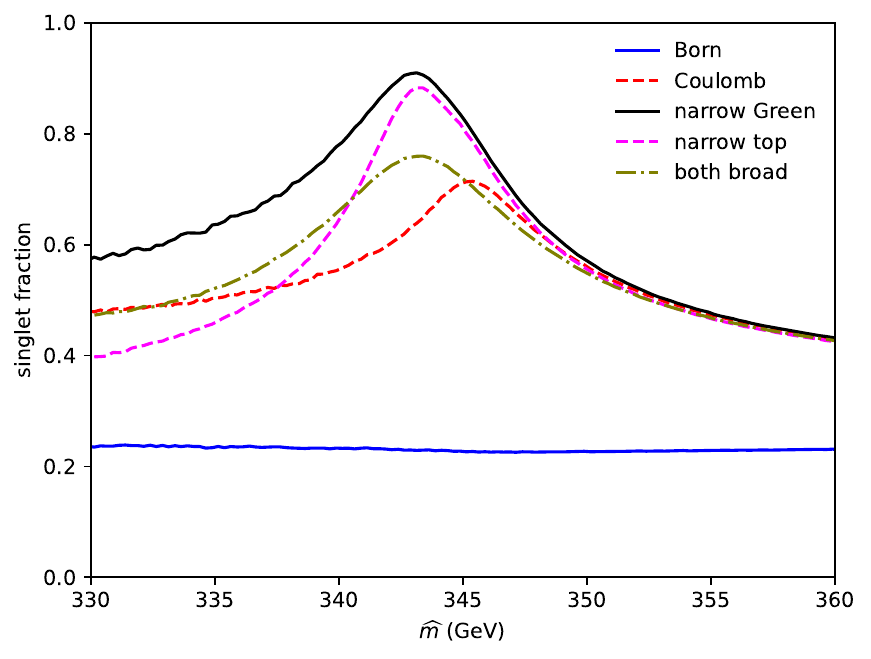}%
\includegraphics[width=0.50\textwidth]{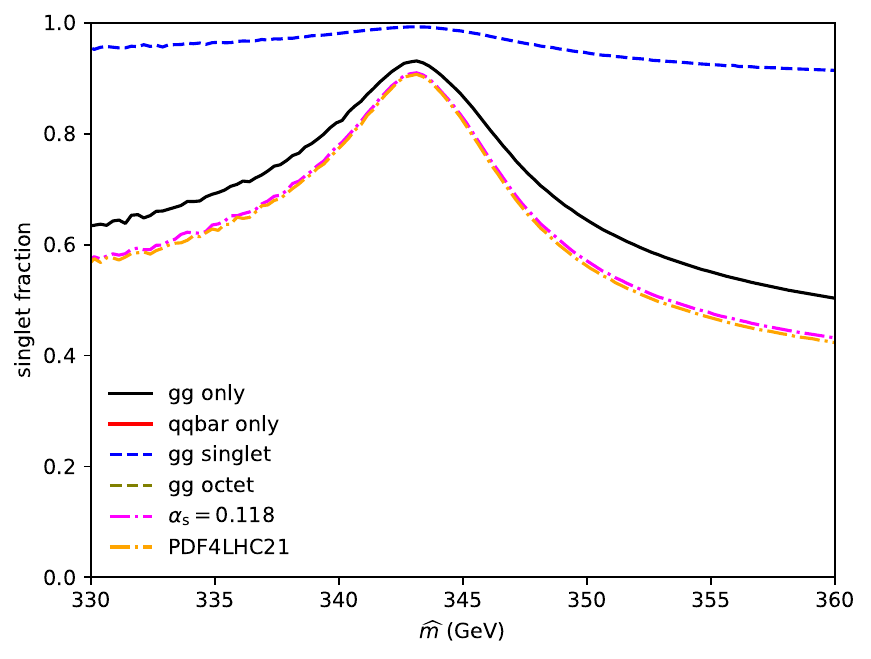}\\[-1mm]
\hspace*{0.25\textwidth}(a)\hspace{0.47\textwidth}(b)
\caption{Colour singlet fraction of $\t\tbar$ production. Note that
the $\q\qbar$-only and $\g\g$-octet options give vanishing singlet
rates.}
\label{fig:singletFrac}
\end{figure}

For top production, the fraction of incoming $\g\g$ in a colour
singlet state is assumed to be $2/7$, and zero for $\q\qbar$.
But then the singlet state is more likely to interact, changing the
singlet fraction, as shown in Fig.~\ref{fig:singletFrac} for the
same model variants as already introduced. The Born option sets
the baseline, while already the Coulomb curve gives a significant
enhancement in the threshold region, further enhanced when also
below-threshold colour-singlet resonance production is included.
It may seem counterintuitive that the singlet fraction drops below
the resonance peak region. The reason is that low-mass top quarks
can give $E >  0$ also for $\mhat < 345$~GeV, and this mechanism
is more competitive below the peaks.
 
\begin{figure}
\includegraphics[width=0.50\textwidth]{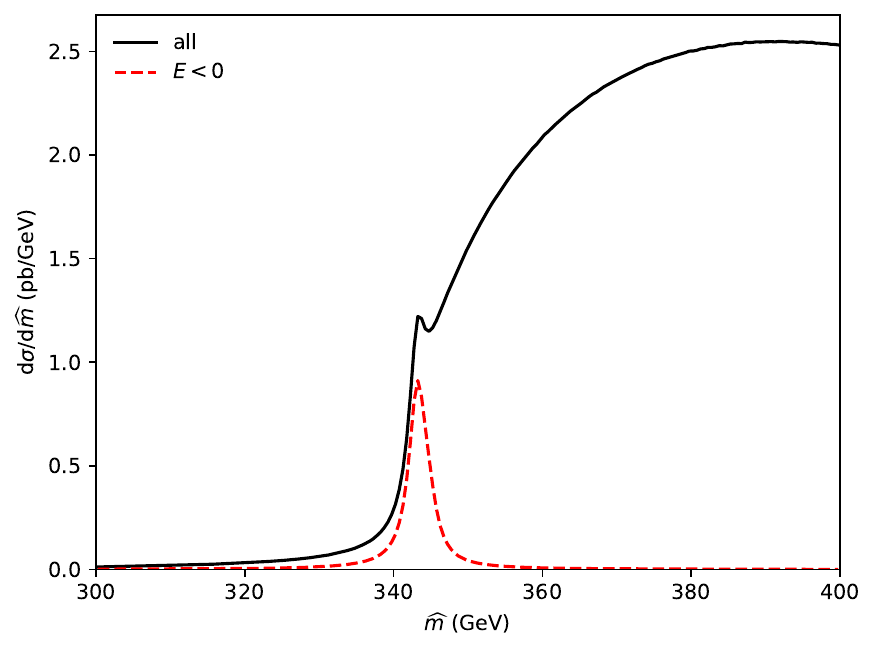}%
\includegraphics[width=0.50\textwidth]{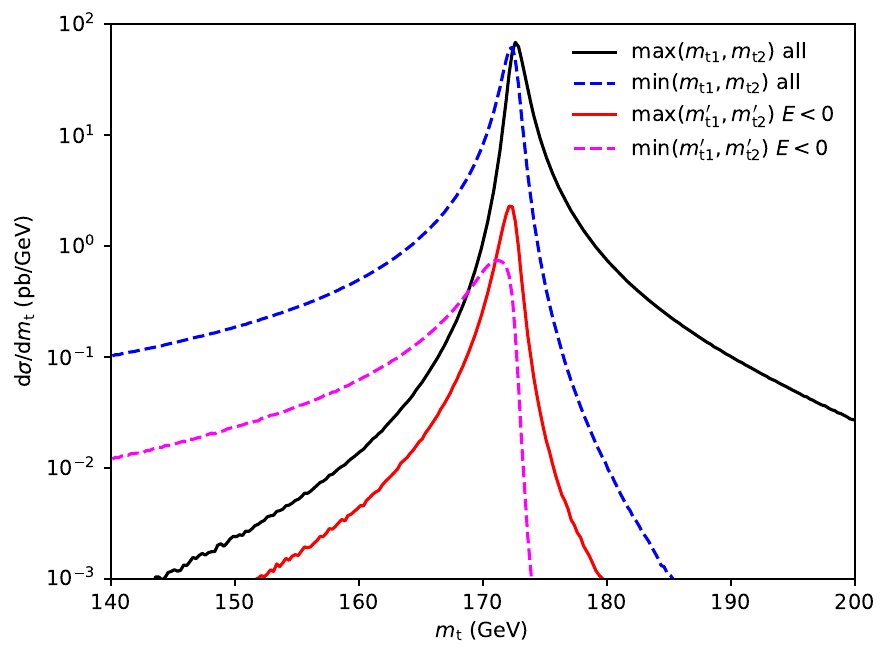}\\[-1mm]
\hspace*{0.25\textwidth}(a)\hspace{0.47\textwidth}(b)
\caption{Narrow-Green scenario (a) $\t$ and $\tbar$ pair mass and
(b) individual masses. For $E < 0$ events the newly selected
$\mtone'$ and  $\mttwo'$ are used, also in the ``all'' sample.}
\label{fig:topMassSel}
\end{figure}

To further study the threshold behaviour, we zoom in on the
300~GeV $< \mhat <$ 400~GeV mass region. This $\mhat$ distribution
is shown in Fig.~\ref{fig:topMassSel}a for the standard narrow-Green
scenario. The non-negligible contribution of $E > 0$ for
$\mhat < 340$~GeV is clearly visible, as is the already mentioned
long tail of $E > 0$ events down to low $\mhat$. The masses of the
individual $\t$ and $\tbar$ quarks are shown in
Fig.~\ref{fig:topMassSel}b. The bulk of these quarks have a fairly
symmetrical distribution around $\mt$, which translates into mirrored
distributions for the maximal and minimal mass of the pair. The masses
for the $E < 0$ subsample, corresponding to the new $\mtone'$ and
$\mttwo'$ choices rather than the original ones, drop more rapidly
for masses above $\mt$, as could be expected, while the below-$\mt$
behaviour is more similar to the rest of the sample.

\begin{figure}[t!]
\includegraphics[width=0.50\textwidth]{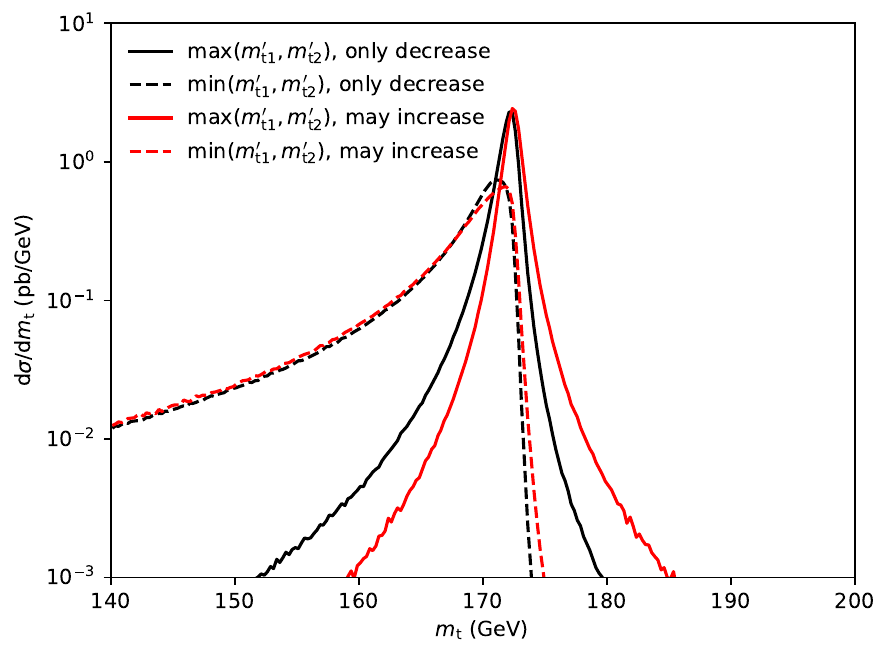}%
\includegraphics[width=0.50\textwidth]{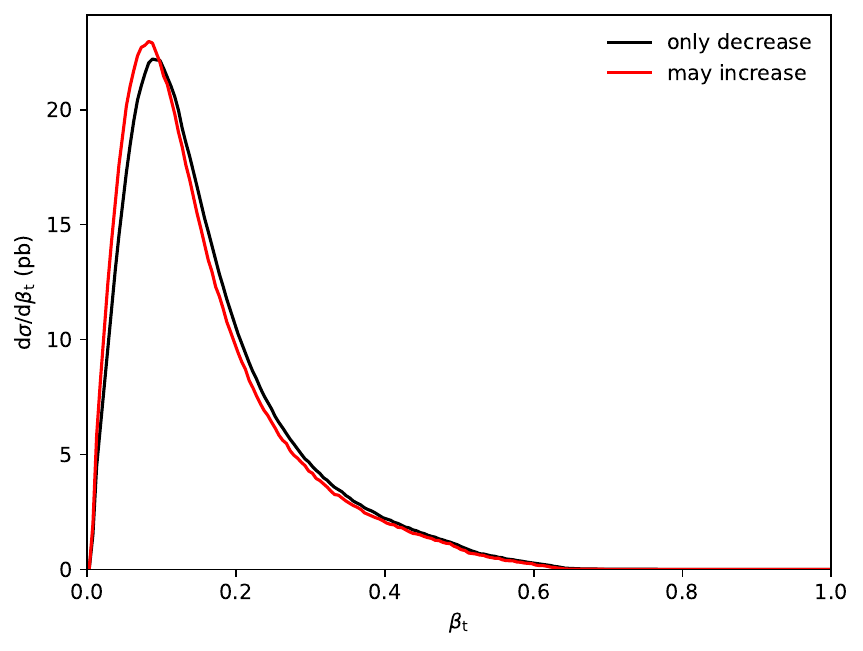}\\[-1mm]
\hspace*{0.25\textwidth}(a)\hspace{0.45\textwidth}(b)
\caption{Below-threshold ($E < 0$) properties: (a) larger and smaller 
of the newly selected $\mtone'$ and $\mttwo'$ masses, and (b) the 
resulting $\betat$, eq.~(\ref{eq:betat}). Comparison of the two
scenarios described in the text.}
\label{fig:topMassAlt}
\end{figure}

This brings us over to another variation that has been considered.
When the original masses $\mtone$ and $\mttwo$ give an $E < 0$
and therefore new $\mtone'$ and $\mttwo'$ are selected, it is
assumed that $\mtone' < \mtone$ and $\mttwo' < \mttwo$. This
is convenient and plausible, but strictly it is only necessary
that $\mtone' + \mttwo' < \mhat < \mtone + \mttwo$, and one
of the two could increase, \eg $\mtone' > \mtone$ at the expense of.
$\mttwo'$. We compare these two options in Fig.~\ref{fig:topMassAlt}.
As can be seen, the smaller of the two new masses are
comparably distributed, whereas unsurprisingly the larger of the two
is more tilted towards larger masses for the alternative handling.
The bulk of the events are located in the peak region, which more
closely agrees, so overall effects are modest. Notably the average
$\betat$ is only moderately smaller. Cross sections for
$E < 0$ only differ at the per cent level; whatever the $\betat'$
in the phase space, the same is used in the $\tilde{G}^{(s,8)}/\betat'$  
weight factor, leaving only smaller differences. For $E > 0$ the two
procedures are identical since no $\mtone'$ and $\mttwo'$ selection
is required, so overall the two options would be hard to separate
experimentally, even under ideal conditions.

In the studies we have aggressively used $\gammatG = 0.02$~GeV so as
to come close to the limit $\gammatG = 0$. To avoid undue
event weights above unity, we have argued for $\gammatG = 0.1$~GeV
as a reasonable choice for everyday use. Differences between
the two are limited. Cross sections for $\gammatG = 0.1$~GeV are
consistent with linear interpolation in Table~\ref{tab:sigmas}
between the narrow-Green $\gammatG = 0.02$~GeV and the narrow-top
$\gammatG = 1.32$~GeV. The $E$ resonance peaks in Fig.~\ref{fig:eandm}
obviously are more smeared-out for a larger $\gammatG$, but the
$\mhat$ spectra are visually indistinguishable, while the $\mt$ ones
in Fig.~\ref{fig:topMassSel}b are only marginally different.
This also suggests that, should one wish to go to the $\gammatG = 0$
limit \eg for the numbers in Table~\ref{tab:sigmas}, linear
extrapolation should be reliable.  
 
\begin{figure}
\includegraphics[width=0.50\textwidth]{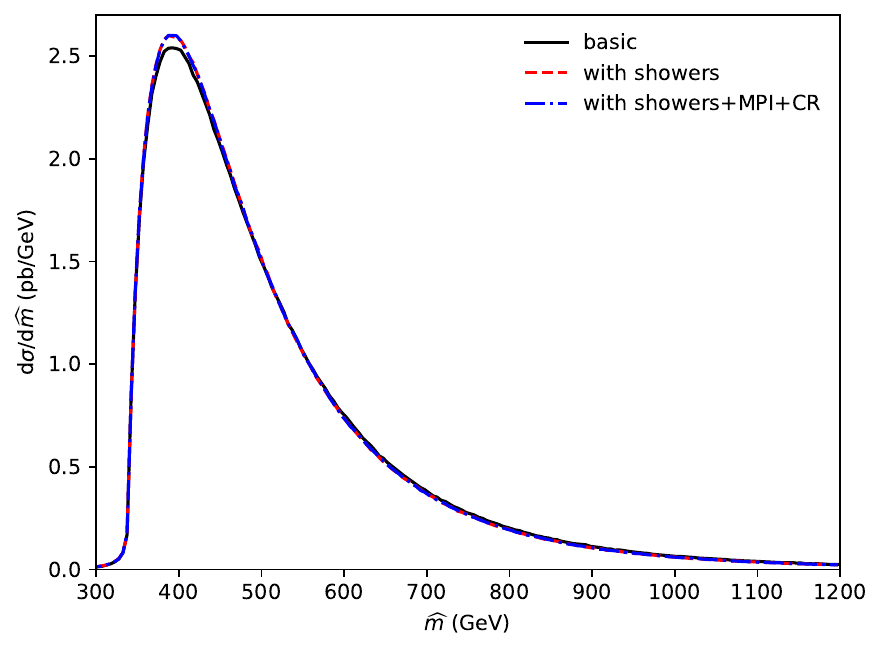}%
\includegraphics[width=0.50\textwidth]{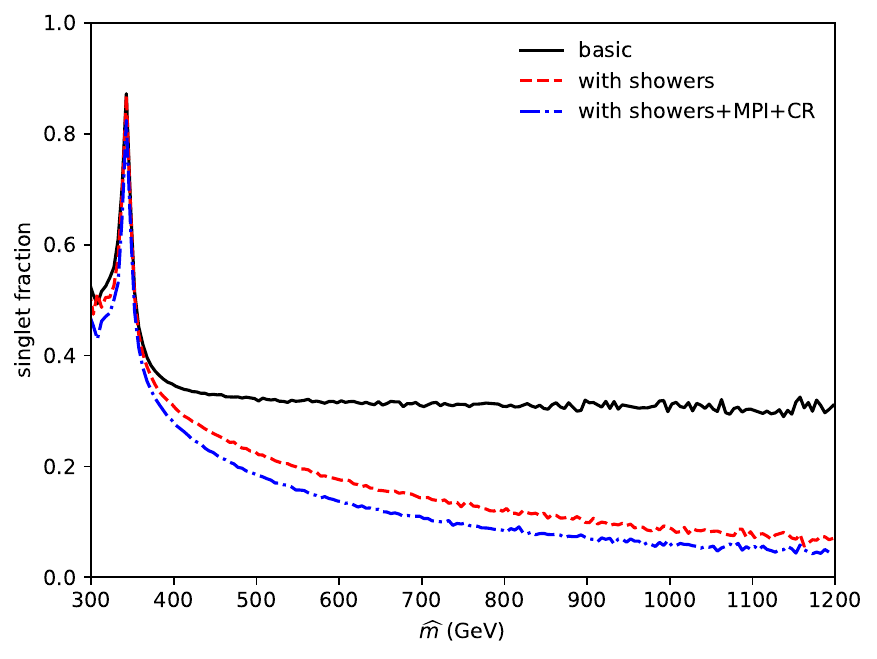}\\[-1mm]
\hspace*{0.25\textwidth}(a)\hspace{0.47\textwidth}(b)
\caption{(a) $\mhat$ distribution and (b) colour singlet fraction for
the basic hard process, with showers added, or also with  multiparton
interactions and colour reconnection. Results are for $10^7$ events,
lower than in other figures owing to the larger generation time for
(almost) complete events than for only the hard process.}
\label{fig:showerResults}
\end{figure}

All the studies so far have dealt with the hard process at the center
of the event. But in real life this is dressed up by initial- and
final-state parton showers, multiparton interactions (MPIs),
colour reconnection (CR) and hadronization. These aspects could change
some of the event properties studied so far. The high top mass  
restricts radiation off the $\t$ and $\tbar$, however, and thereby
any shift of the $\t\tbar$ invariant mass $\mhat$, up or down.
Furthermore the default initial-state showers, which do involve lighter
partons and therefore can radiate more, shift the whole
$\t\tbar$ system as a unit, with unchanged $\mhat$. MPIs and CR do not
affect the kinematics. So overall $\mhat$ is almost unchanged,
Fig.~\ref{fig:showerResults}a, with only a small shift from the  
high-mass tail to the peak region. More visible is the reduced
colour-singlet fraction, Fig.~\ref{fig:showerResults}b. The emission of a
gluon from either of $\t$ and $\tbar$ will turn a singlet into an octet,
and the phase space for such emissions goes up with increasing $\mhat$.
On the other hand, the relevance of $\t$ and $\tbar$ being in a singlet
state drops with $\mhat$. Furthermore, a warning is that results are
sensitive to the lower stopping scale of shower activity.
The MPIs add an underlying partonic activity that then the CR scheme
can exploit to turn singlets into octets, or occasionally the other
way around, and again a larger original $\t\tbar$ separation leaves
more room for such reconnections. But also here a warning is in place,
that the reconnection criteria in CR models have not been developed
with the role of top quarks in mind, so are open for further refinements.
Results here should therefore be seen as qualitative rather than
quantitative.    

\section{Electroweak production studies}\label{sec:electroweakproduction}

In addition to the $\g\g \to \t \tbar$ and $\q\qbar \to \g^* \to \t\tbar$
QCD production processes, a third possibility for hadron colliders is the
$\q\qbar \to \gamma^*/\Z^0 \to \t \tbar$ electroweak process. Its cross
section is down by two orders of magnitude relative to the QCD $\q\qbar$
one, however, and by three orders relative to $\g\g$. To give some numbers,
to be compared with Table~\ref{tab:sigmas}, the total cross section
is 0.40~pb at the Born level, which increases to 0.62~pb in the
narrow-Green scenario. Corresponding numbers for $\mhat < 345$~GeV are
0.002~pb and 0.021~pb, respectively. The reason for the factor-of-ten 
ratio of the last two numbers obviously is that this channel is pure
colour singlet, so that the Coulomb/Green's function enhancement is 
maximized. In Table~\ref{tab:sigmas} the ratio between the $\g\g$ 
singlet + $\q\qbar$ octet and the Born numbers is also of order ten.

\begin{figure}
\includegraphics[width=0.50\textwidth]{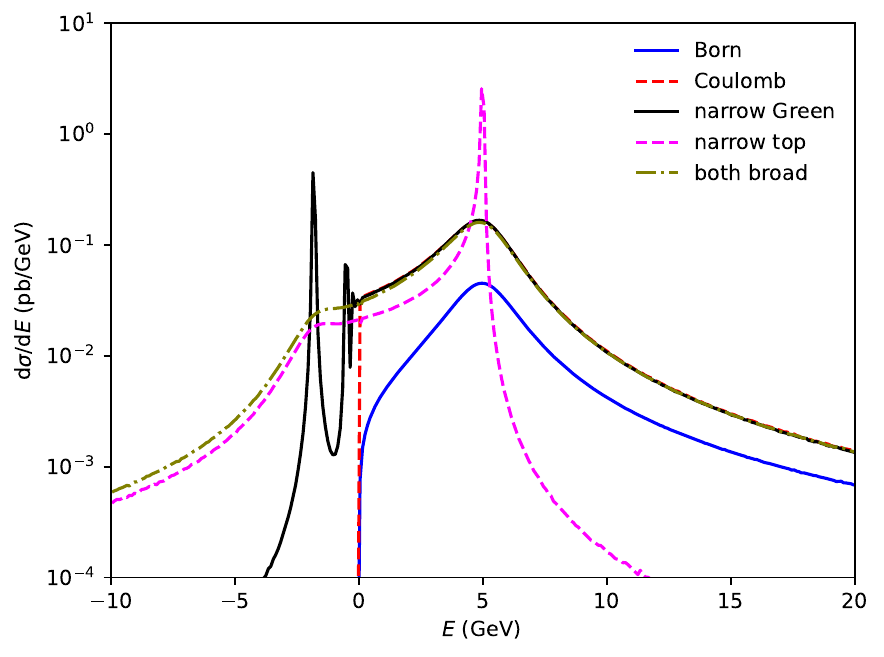}%
\includegraphics[width=0.50\textwidth]{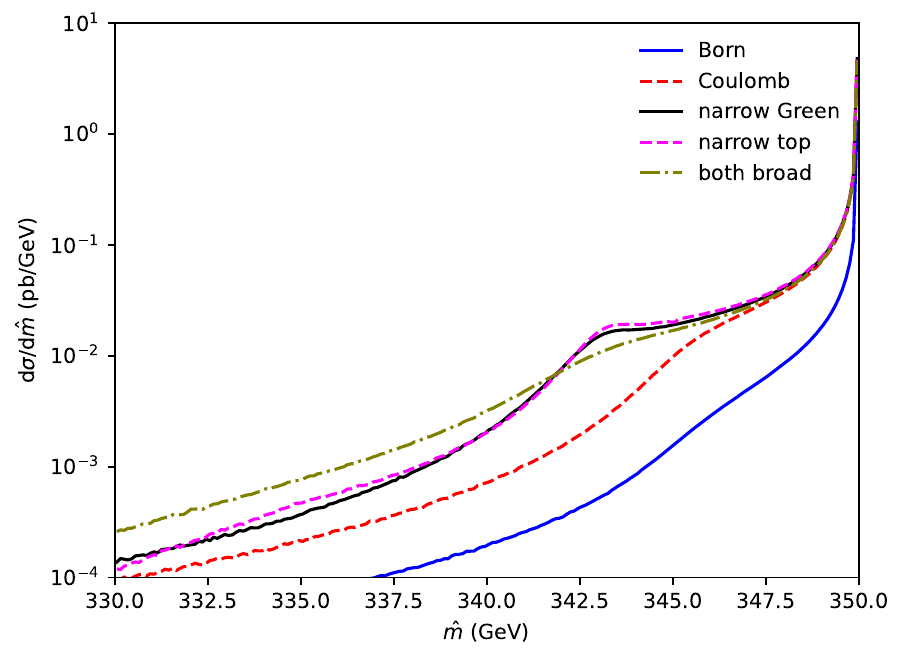}\\[-1mm]
\hspace*{0.25\textwidth}(a)\hspace{0.47\textwidth}(b)
\caption{Two distributions for a 350~GeV $\e^+\e^-$ collider with
bremsstrahlung included, (a) threshold energy $E$ and (b) invariant
$\t\tbar$ mass $\mhat$. In (a) the Coulomb, narrow-Green and both-broad
curves overlap for $E > 0$.}
\label{fig:eedist}
\end{figure}

Given the small electroweak impact at the LHC, it is more 
interesting to go ``back to the roots'', namely to the 
$\e^+\e^- \to \gamma^*/\Z^0 \to \t \tbar$ process, where the top threshold
studies begun. Now having the formalism in place, it is straightforward
to apply it to this case. As example, key distributions for an
$\Ecm = 350$~GeV collider are shown in Fig.~\ref{fig:eedist},
comparing the same five models as for LHC. The universal QED
bremsstrahlung is included here, by the introduction of leptonic PDFs
$f_{\e}^{\e}(x, Q^2)$, such that $\mhat = \sqrt{x_1x_2}\Ecm < \Ecm$.  
A further $\mhat$ reduction could come from beamstrahlung, but this is
machine-dependent and is not considered here. The plots reflect the
roles of two key effects. On the one hand the bremsstrahlung spectrum
peaks at low photon and high electron energies, $x \approx 1$, thereby
favouring $\mhat \approx \Ecm$ and $E \approx \Ecm - 2\mt \approx 5$~GeV,
as especially visible in the narrow-top model. On the other hand the
below-threshold contributions, especially visible in the $E$ spectrum
of the narrow-Green main scenario, give an enhancement which peaks 
slightly below $\mhat \approx 2\mt - 2~\mathrm{GeV} \approx 343$~GeV.

\begin{figure}
\includegraphics[width=0.50\textwidth]{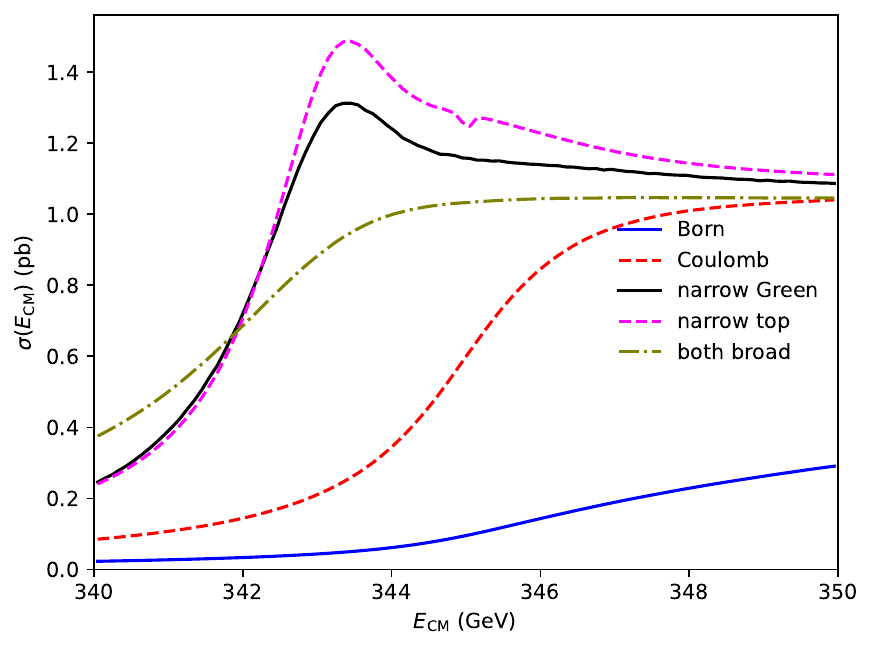}%
\includegraphics[width=0.50\textwidth]{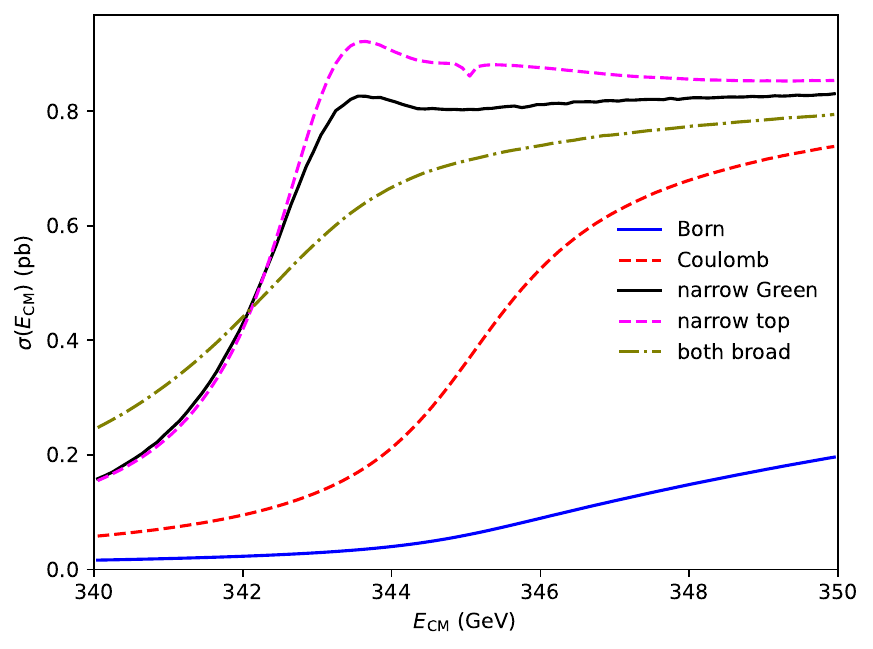}\\[-1mm]
\hspace*{0.25\textwidth}(a)\hspace{0.47\textwidth}(b)
\caption{The $\t\tbar$ cross section in $\e^+\e^-$ collisions as a 
function of the collision energy $\Ecm$, in (a) without QED 
initial-state bremsstrahlung and in (b) with it.}
\label{fig:eesigma}
\end{figure}

While the $\mhat$ distribution at a fixed $\Ecm$ could be
used to pin down the top mass, it would be dependent on the details
of the event reconstruction, so a threshold scan of the total $\t\tbar$
cross section usually is advocated as a cleaner measurement.
Fig.~\ref{fig:eesigma} shows the energy dependence of this
cross section, without and with bremsstrahlung. Without it,
the cross section enhancement approximately 2~GeV below the $\t\tbar$
threshold is clearly visible, but this peak almost completely
disappears when bremsstrahlung is included, since part of the
$\e^+\e^-$ flux given by the $f_{\e}^{\e}(x, Q^2)$ factors is shifted
to below the resonance region, where the cross section is reduced.
Actually, not only is the peak smeared away by bremsstrahlung, 
but the cross section is reduced over the whole energy range studied. 
The reason is that, although the $\e^+\e^-$ flux peaks at $\mhat = \Ecm$, 
there is a long and non-negligible tail to $\mhat$ values well below 
the narrow mass range of this figure, and that tail is all but lost 
for $\t\tbar$ production. What remains as a distinguishing feature, 
between the Coulomb and narrow-Green scenarios, is that the rise of
the cross section is steeper in the latter case. 

As an aside, the small dip for the narrow-top curve at 345~GeV
is caused by some hardcoded threshold-energy numerical safety margins 
in the current code. It could be addressed by recompiling with a 
reduced safety, at the risk of problems for other processes.
Since the narrow-top scenario is not our preferred one anyway,
for now we leave it be. 

\begin{figure}
\hspace*{0.25\textwidth}%
\includegraphics[width=0.50\textwidth]{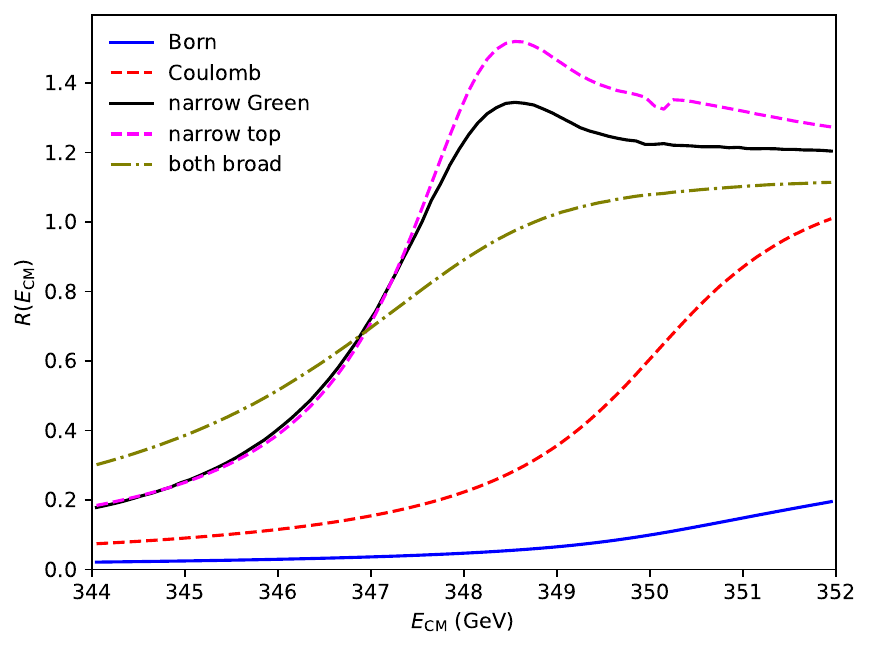}
\caption{Ratio $R$ of the QED $\t\tbar$ to $\mu^+\mu^-$ cross sections
in $\e^+\e^-$ collisions, as further described in the text.}
\label{fig:eesigmaHoang}
\end{figure}

The article \cite{Hoang:2000yr} compares four different NNLO 
calculations of the threshold-energy behaviour. A broad agreement is 
shown, but with non-negligible differences both between the four
and within each as scale choices are varied. As a cross-check, in 
Fig.~\ref{fig:eesigmaHoang} we plot results in the same $\Ecm$ range 
as in Fig.~1 of that article, with $\mt = 175.05$~GeV and 
$\gammat = 1.43$~GeV. In it, results are restricted to the pure QED
$\gamma^*$ exchange. This is a not unreasonable starting point, 
for two reasons. Firstly the vector Born cross section has the
familiar $\betat$ threshold factor, but the axial-vector part has a 
$\betat^3$ one, suppressing it in the region studied. Secondly the
$v_{\e}$ electron-to-$\Z^0$ vector coupling is small. Furthermore,
no bremsstrahlung effects are included. Finally, results are 
presented for $R = \sigma_{\t\tbar} / \sigma_{\mathrm{pt}}$, where 
$\sigma_{\mathrm{pt}} = 4\pi\aem^2/3s$ is the Born-level
$\e^+\e^- \to \gamma^* \to \mu^+\mu^-$ cross section. By coincidence 
$\sigma_{\mathrm{pt}}$ is of order 1~pb, which makes the vertical
scales of Fig.~\ref{fig:eesigma}a and Fig.~\ref{fig:eesigmaHoang} 
appear so similar, although dimensionally different.      

Overall the agreement between the narrow-Green model and the four
scenarios in  \cite{Hoang:2000yr} is good, both in terms of general
shape and cross sections below, on and above the peak. There is 
a visible difference between the peak position, which is around
348~GeV for us and around or below 347~GeV there. But one should 
note the non-negligible scale dependence, and also ambiguities in
the top mass definition, so overall agreement is quite encouraging.

\section{Decay studies}\label{sec:decaystudies}

\begin{figure}
\includegraphics[width=0.50\textwidth]{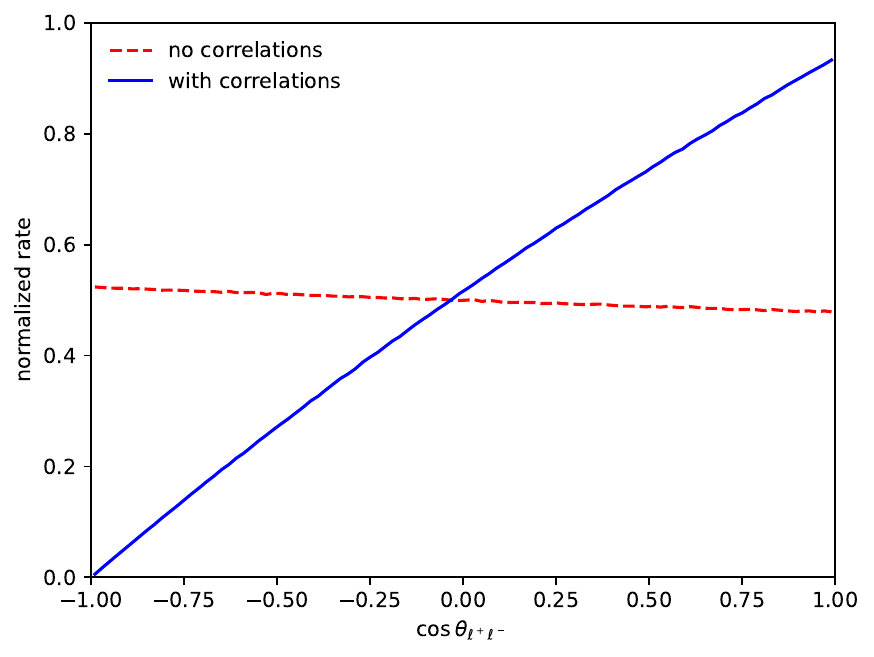}%
\includegraphics[width=0.50\textwidth]{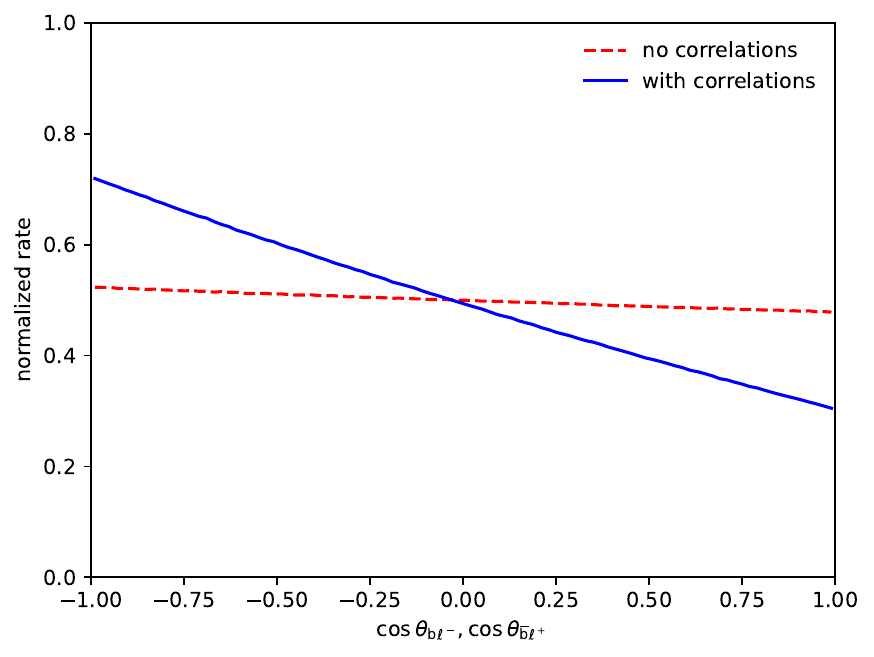}\\[-1mm]
\hspace*{0.25\textwidth}(a)\hspace{0.45\textwidth}(b)\\[2mm]
\includegraphics[width=0.50\textwidth]{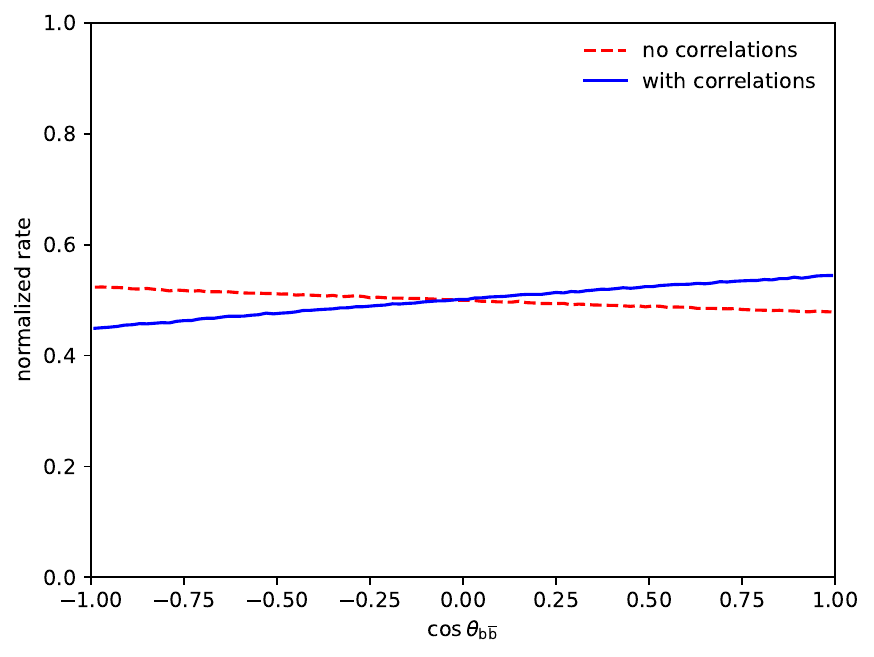}%
\includegraphics[width=0.50\textwidth]{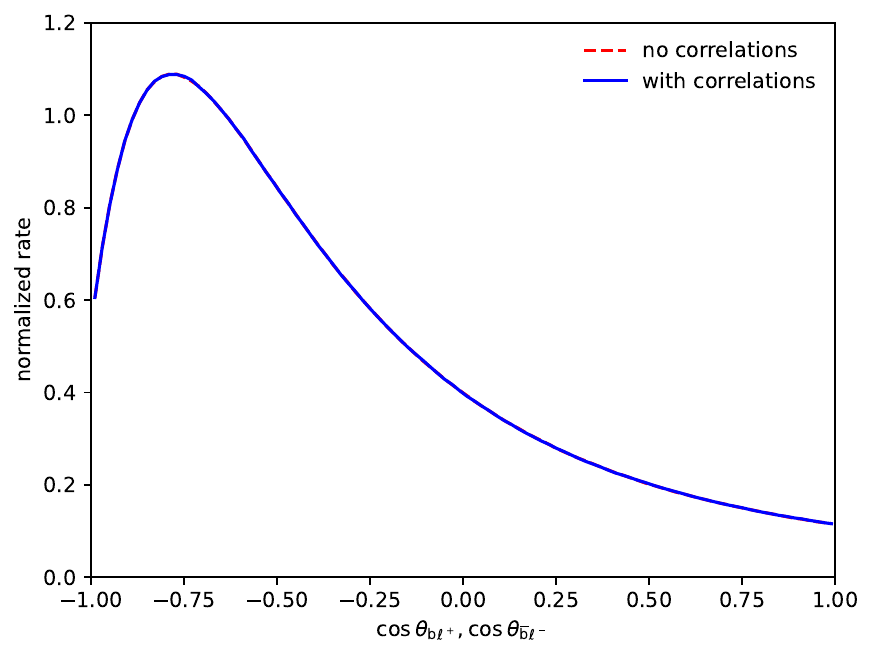}\\[-1mm]
\hspace*{0.25\textwidth}(c)\hspace{0.47\textwidth}(d)
\caption{Angular correlations in near-threshold
pseudoscalar $\to \t\tbar \to \b\W^+\bbar\W^- \to \b\ell^+\nu_{\ell}%
\bbar\overline{\nu}_{ell}$ decay sequences, without or with spin
correlations between the $\t$ and $\tbar$. In frame (d) the two
curves overlap perfectly.}
\label{fig:decayAngles}
\end{figure}

One of the key tools in the experimental isolation of a signal
is the pseudoscalar nature of $\t\tbar$ states near threshold,
for the dominant $\g\g \to \t\tbar$ process,
whether from pseudo-bound below-threshold states or not. This leads
to angular correlations between the decay products. Notably
for the $\ell^+\ell^-$ ($\ell$ a lepton) pair in the
$\t\tbar \to \b\W^+\bbar\W^- \to \b\ell^+\nu_{\ell}\bbar\overline{\nu}_{\ell}$
decay sequence, which plays an important role in the experimental
analysis of candidate events.

Traditionally \Pythia does not include angular correlations between
the $\t$ and $\tbar$ decay chains, which is a good approximation
well above threshold, although more detailed knowledge may be required
for certain experimental measurements. Inside each decay, a
matrix-element weight $(p_{\t}p_{\ell})(p_{\nu_{\ell}}p_{\b})$ is applied
to the $\W$ decay angles.

As a first step to improve this situation, we have implemented the
matrix element for the decay of a pseudoscalar initial state via
$\t\tbar$ to the six-body final state. The expression has been
calculated using \textsc{FeynCalc}~9.3
\cite{Mertig:1990an,Shtabovenko:2020gxv} and simplified using
\textsc{Form}~4.2 \cite{Ruijl:2017dtg}. The calculation has been
performed under the assumption of a pseudoscalar Yukawa-like coupling
between the pseudoscalar initial state and the top quarks with full
off-shell electroweak decays $\t \to \W\b$ and $\W \to \ell\nu_{\ell}$.
All leptons and the bottom quark have been treated as massless. The
numerical value of the coupling strength is irrelevant here, as this
expression is only used to reweight uncorrelated $\t/\tbar$ and $\W^{\pm}$
decays, isotropic in their respective rest frames, to the fully correlated
angular distributions. We have cross checked the final expression
against a custom \textsc{MadGraph5}\_aMC@NLO \cite{Alwall:2014hca}
amplitude obtained within its MSSM model, finding agreement to
machine-level precision. A limitation is that this calculation
does not include transition from correlated to independent
$\t$ and $\tbar$ decays --- this would be better handled by dedicated
matrix elements in the above-threshold region. Relevant expressions
could either be implemented in \textsc{Pythia} in the future or
interfaced to it via dedicated matrix-element providers. A simple
interim option exists to set the beginning and end of a linear
transition region between the two.

For a simple study, we consider the threshold region, $E < 10$~GeV,
where purely pseudoscalar distributions can be assumed. Events are
studied in the rest frame of the $\t\tbar$ pair. Resulting angles
between leptons and $b$ quarks are shown in Fig.~\ref{fig:decayAngles}.
The most spectacular is the $\ell^+\ell^-$ one. If additionally the
$\t$ and $\tbar$ had been boosted along the $\t\tbar$ axis to their
respective rest frame it would have given a straight line 
\cite{Fuks:2021xje,Fuks:2024yjj}
\begin{equation}
\frac{1}{\sigma} \, \frac{\d\sigma}{d\cos\theta_{\ell^+\ell^-}} =  
\frac{1 + \cos\theta_{\ell^+\ell^-}}{2}~.
\end{equation}
As shown here, the motion of the $\t$ and $\tbar$ apart from each other
gives a slight bias towards small $\cos\theta_{\ell^+\ell^-}$, as is
also seen from the tilt of the no-correlations curve. Correlations 
are also visible for the $\b\ell^-/\bbar\ell^+$ and $\b\bbar$ angles,
but much smaller. The $\b\ell^+/\bbar\ell^-$ correlate two fermions
from the same decay chain, and it is a good sanity check that the
distribution here coincides with the simple uncorrelated $\t$ and
$\tbar$ case.

Other less transparent angles have been proposed and used, but are not
studied here.

\section{Summary and conclusions}\label{sec:summary}

The early articles by Fadin and Khoze offered the first
applications of a Green's function formalism to the formation of
pseudo-bound ``toponium'' states below threshold and cross-section
enhancements above it. So far, not all the details of the
calculations have been published, and relevant notes have since been
lost. When interest arose to revive this formalism for comparisons
with new CMS and ATLAS data on the threshold cross section,
a first attempt therefore came to double-count the effects of
top mass smearing. The main intent of this article is to set the
record straight, and understand the origin of the FK
Green's functions. A key result is eq.~(\ref{eq:greensdelta}),
that in the $\gammat \to 0$ limit the Green's functions reduce
to the Coulomb factors above threshold, plus a set of delta functions
below threshold for the singlet part. Another is the realization 
that eq.~(\ref{eq:trick}) allows each 
Coulomb factor to be written in a form that has the same
pole structure as the below-threshold states. This allows a
convolution with non-relativistic Breit--Wigners for the $\t$ and 
$\tbar$ by calculus of residues. In the resulting expressions
the $\gammat = 0$ origin is no longer transparent. It is further 
masked by a choice of the $E$ energy scale that is an approximation 
of the correct Green's function argument.    

Having traced the origins, we can now more confidently move on with
the simulation of the top threshold behaviour. Specifically, instead
of relying on an inclusively smeared $\t\tbar$ mass spectrum, we can 
correlate the event-by-event $\t$ and $\tbar$ masses with the intended
threshold behaviour, \ie be differential in $(\mhat, \mtone, \mttwo)$
instead of in $\mhat$ only. For technical reasons a small width is still
kept in the Green's function expression, but at a level where predictions
closely correlate with the $\gammat \to 0$ limit of the Green's
functions. Some key numbers are that the contribution from the
below-threshold pseudo-bound states is $\sim 4.4$~pb, but that the
Breit--Wigner smearing gives $\sim 6.5$~pb for $\mhat < 2 \mt$.
These numbers are for default assumptions about the colour singlet
fraction, $\as$ scales, PDF selection, and much more, so there is
some leeway in them. But, as importantly, the Coulomb/Green's function
enhancement above threshold in itself affects any experimental
study with current or foreseeable mass resolutions in hadron
colliders.

Our aim is that the Monte Carlo implementation of FK
formulae will be useful in the continued study of the $\t\tbar$
threshold behaviour. Yet there is every reason to be modest.
The original calculation 35+ years ago defined the state of the art
at the time, but advances have been made since. The baseline,
above which the ``toponium'' signal should stand out, now is defined
by higher-order perturbative calculations rather than the resummed
Coulomb expression, see e.g. \cite{Beneke:2024sfa, Nason:2025hix}. 
Green's functions have been derived with more
sophisticated QCD potentials, where numerical rather than analytical
solutions have been obtained, and these have been provided to the
experimental community. Further, for experimental studies the top
decay angular correlations are crucial, and when they now are provided
in \Pythia it is only for the threshold region. Some of these issues
could resolve themselves naturally; \eg a transition to
matrix-element-based calculations some small distance above the
threshold would bring with it the desired angular correlations. 

What remains it that we now can offer a simple but better-defined
and better-understood model, with extensive flexibility to explore
plausible variations. 

\section*{Acknowledgements}

We are very grateful to Victor Fadin for numerous contributions to and
discussions on the topics of this article.
VAK thanks Alexandre Rozanov for useful discussions.
The work of CTP is supported by the Deutsche Forschungsgemeinschaft
(DFG) under grant 396021762 - TRR 257: Particle Physics Phenomenology
after the Higgs Discovery.

\bibliographystyle{utphys}
\bibliography{topRevisit}

\providecommand{\href}[2]{#2}\begingroup\raggedright\begin{thebibliography}{10}

\bibitem{Kobayashi:1973fv}
M.~Kobayashi and T.~Maskawa, ``{CP Violation in the Renormalizable Theory of
  Weak Interaction},'' {\em Prog. Theor. Phys.} {\bf 49} (1973) 652--657.

\bibitem{Perl:1975bf}
M.~L. Perl {\em et.~al.}, ``{Evidence for Anomalous Lepton Production in e+ -
  e- Annihilation},'' {\em Phys. Rev. Lett.} {\bf 35} (1975) 1489--1492.

\bibitem{E288:1977xhf}
{\bf E288} Collaboration, S.~W. Herb {\em et.~al.}, ``{Observation of a Dimuon
  Resonance at 9.5 GeV in 400 GeV Proton-Nucleus Collisions},'' {\em Phys. Rev.
  Lett.} {\bf 39} (1977) 252--255.

\bibitem{PLUTO:1978jrw}
{\bf PLUTO} Collaboration, C.~Berger {\em et.~al.}, ``{Jet Analysis of the
  $\Upsilon$ (9.46) Decay Into Charged Hadrons},'' {\em Phys. Lett. B} {\bf 82}
  (1979) 449--455.

\bibitem{ARGUS:1987xtv}
{\bf ARGUS} Collaboration, H.~Albrecht {\em et.~al.}, ``{Observation of B0 -
  anti-B0 Mixing},'' {\em Phys. Lett. B} {\bf 192} (1987) 245--252.

\bibitem{Ali:1987jv}
A.~Ali, ``{B0 - anti-B0 MIXINGS: A REAPPRAISAL},'' in {\em {First International
  Symposium on the 4th Family of Quarks and Leptons}}, 7, 1987.

\bibitem{Blinov:1988cc}
A.~E. Blinov, V.~A. Khoze, and N.~G. Uraltsev, ``{Physics of Top and {CP}
  Violation in $B$ Decays in the Light of the Argus Measurements},'' {\em Int.
  J. Mod. Phys. A} {\bf 4} (1989) 1933.

\bibitem{LEP:1991hsu}
{\bf LEP, ALEPH, DELPHI, L3, OPAL} Collaboration, G.~Alexander {\em et.~al.},
  ``{Electroweak parameters of the $Z^0$ resonance and the Standard Model: the
  LEP Collaborations},'' {\em Phys. Lett. B} {\bf 276} (1992) 247--253.

\bibitem{CDF:1994vkk}
{\bf CDF} Collaboration, F.~Abe {\em et.~al.}, ``{Evidence for top quark
  production in $\bar{p}p$ collisions at $\sqrt{s} = 1.8$ TeV},'' {\em Phys.
  Rev. D} {\bf 50} (1994) 2966--3026.

\bibitem{CDF:1995wbb}
{\bf CDF} Collaboration, F.~Abe {\em et.~al.}, ``{Observation of top quark
  production in $\bar{p}p$ collisions},'' {\em Phys. Rev. Lett.} {\bf 74}
  (1995) 2626--2631, \href{http://xxx.lanl.gov/abs/hep-ex/9503002}{{\tt
  hep-ex/9503002}}.

\bibitem{D0:1995jca}
{\bf D0} Collaboration, S.~Abachi {\em et.~al.}, ``{Observation of the top
  quark},'' {\em Phys. Rev. Lett.} {\bf 74} (1995) 2632--2637,
  \href{http://xxx.lanl.gov/abs/hep-ex/9503003}{{\tt hep-ex/9503003}}.

\bibitem{Bigi:1986jk}
I.~I.~Y. Bigi, Y.~L. Dokshitzer, V.~A. Khoze, J.~H. Kuhn, and P.~M. Zerwas,
  ``{Production and Decay Properties of Ultraheavy Quarks},'' {\em Phys. Lett.
  B} {\bf 181} (1986) 157--163.

\bibitem{Fadin:1987wz}
V.~S. Fadin and V.~A. Khoze, ``{Threshold Behavior of Heavy Top Production in
  e+ e- Collisions},'' {\em JETP Lett.} {\bf 46} (1987) 525--529.

\bibitem{Fadin:1988fn}
V.~S. Fadin and V.~A. Khoze, ``{Production of a pair of heavy quarks in e+ e-
  annihilation in the threshold region},'' {\em Sov. J. Nucl. Phys.} {\bf 48}
  (1988) 309--313.

\bibitem{Fadin:1989fd}
V.~S. Fadin, V.~A. Khoze, and T.~Sj{\"o}strand, ``{ON THE THRESHOLD BEHAVIOR OF
  HEAVY TOP PRODUCTION},'' in {\em {24th Rencontres de Moriond: New Results in
  Hadronic Interactions}}, pp.~19--32, 1989.

\bibitem{Fadin:1990wx}
V.~S. Fadin, V.~A. Khoze, and T.~Sj{\"o}strand, ``{On the Threshold Behavior of
  Heavy Top Production},'' {\em Z. Phys. C} {\bf 48} (1990) 613--622.

\bibitem{Fadin:1991zw}
V.~S. Fadin and V.~A. Khoze, ``{Production of a pair of $t \bar{t}$ quarks near
  threshold},'' {\em Sov. J. Nucl. Phys.} {\bf 53} (1991) 692--698.

\bibitem{Kwong:1990iy}
W.-k. Kwong, ``{Threshold production of $t \bar{t}$ pairs by $e^{+} e^{-}$
  collisions},'' {\em Phys. Rev. D} {\bf 43} (1991) 1488--1499.

\bibitem{Strassler:1990nw}
M.~J. Strassler and M.~E. Peskin, ``{The Heavy top quark threshold: QCD and the
  Higgs},'' {\em Phys. Rev. D} {\bf 43} (1991) 1500--1514.

\bibitem{Hoang:2000yr}
A.~H. Hoang {\em et.~al.}, ``{Top - anti-top pair production close to
  threshold: Synopsis of recent NNLO results},'' {\em Eur. Phys. J. direct}
  {\bf 2} (2000), no.~1 3, \href{http://xxx.lanl.gov/abs/hep-ph/0001286}{{\tt
  hep-ph/0001286}}.

\bibitem{Hagiwara:2008df}
K.~Hagiwara, Y.~Sumino, and H.~Yokoya, ``{Bound-state Effects on Top Quark
  Production at Hadron Colliders},'' {\em Phys. Lett. B} {\bf 666} (2008)
  71--76, \href{http://xxx.lanl.gov/abs/0804.1014}{{\tt 0804.1014}}.

\bibitem{Kiyo:2008bv}
Y.~Kiyo, J.~H. Kuhn, S.~Moch, M.~Steinhauser, and P.~Uwer, ``{Top-quark pair
  production near threshold at LHC},'' {\em Eur. Phys. J. C} {\bf 60} (2009)
  375--386, \href{http://xxx.lanl.gov/abs/0812.0919}{{\tt 0812.0919}}.

\bibitem{Sumino:2010bv}
Y.~Sumino and H.~Yokoya, ``{Bound-state effects on kinematical distributions of
  top quarks at hadron colliders},'' {\em JHEP} {\bf 09} (2010) 034,
  \href{http://xxx.lanl.gov/abs/1007.0075}{{\tt 1007.0075}}. [Erratum: JHEP 06,
  037 (2016)].

\bibitem{Ju:2020otc}
W.-L. Ju, G.~Wang, X.~Wang, X.~Xu, Y.~Xu, and L.~L. Yang, ``{Top quark pair
  production near threshold: single/double distributions and mass
  determination},'' {\em JHEP} {\bf 06} (2020) 158,
  \href{http://xxx.lanl.gov/abs/2004.03088}{{\tt 2004.03088}}.

\bibitem{Nason:1987xz}
P.~Nason, S.~Dawson, and R.~K. Ellis, ``{The Total Cross-Section for the
  Production of Heavy Quarks in Hadronic Collisions},'' {\em Nucl. Phys. B}
  {\bf 303} (1988) 607--633.

\bibitem{CMS:2025kzt}
{\bf CMS} Collaboration, A.~Hayrapetyan {\em et.~al.}, ``{Observation of a
  pseudoscalar excess at the top quark pair production threshold},'' {\em Rept.
  Prog. Phys.} {\bf 88} (2025), no.~8 087801,
  \href{http://xxx.lanl.gov/abs/2503.22382}{{\tt 2503.22382}}.

\bibitem{ATLAS:2026dbe}
{\bf ATLAS} Collaboration, G.~Aad {\em et.~al.}, ``{Observation of a
  cross-section enhancement near the $t\bar{t}$ production threshold in
  $\sqrt{s}=13$ TeV $pp$ collisions with the ATLAS detector},''
  \href{http://xxx.lanl.gov/abs/2601.11780}{{\tt 2601.11780}}.

\bibitem{CMS-PAS-TOP-25-002}
{\bf CMS} Collaboration, ``{Observation of a pseudoscalar excess at the top
  quark pair production threshold in the single lepton channel},'' tech. rep.,
  CERN, Geneva, 2026.

\bibitem{Fuks:2021xje}
B.~Fuks, K.~Hagiwara, K.~Ma, and Y.-J. Zheng, ``{Signatures of toponium
  formation in LHC run 2 data},'' {\em Phys. Rev. D} {\bf 104} (2021), no.~3
  034023, \href{http://xxx.lanl.gov/abs/2102.11281}{{\tt 2102.11281}}.

\bibitem{Maltoni:2024tul}
F.~Maltoni, C.~Severi, S.~Tentori, and E.~Vryonidou, ``{Quantum detection of
  new physics in top-quark pair production at the LHC},'' {\em JHEP} {\bf 03}
  (2024) 099, \href{http://xxx.lanl.gov/abs/2401.08751}{{\tt 2401.08751}}.

\bibitem{Alwall:2014hca}
J.~Alwall, R.~Frederix, S.~Frixione, V.~Hirschi, F.~Maltoni, O.~Mattelaer,
  H.~S. Shao, T.~Stelzer, P.~Torrielli, and M.~Zaro, ``{The automated
  computation of tree-level and next-to-leading order differential cross
  sections, and their matching to parton shower simulations},'' {\em JHEP} {\bf
  07} (2014) 079, \href{http://xxx.lanl.gov/abs/1405.0301}{{\tt 1405.0301}}.

\bibitem{Fuks:2024yjj}
B.~Fuks, K.~Hagiwara, K.~Ma, and Y.-J. Zheng, ``{Simulating toponium formation
  signals at the LHC},'' {\em Eur. Phys. J. C} {\bf 85} (2025), no.~2 157,
  \href{http://xxx.lanl.gov/abs/2411.18962}{{\tt 2411.18962}}.

\bibitem{Fuks:2025wtq}
B.~Fuks, K.~Hagiwara, K.~Ma, L.~Munoz-Aillaud, and Y.-J. Zheng, ``{Prospects
  for toponium formation at the LHC in the single-lepton mode},''
  \href{http://xxx.lanl.gov/abs/2509.03596}{{\tt 2509.03596}}.

\bibitem{Fuks:2025toq}
B.~Fuks, A.~Hossain, and J.~Keaveney, ``{Statistical indications of toponium
  formation in top quark pair production},'' {\em Phys. Lett. B} {\bf 873}
  (2026) 140179, \href{http://xxx.lanl.gov/abs/2511.02040}{{\tt 2511.02040}}.

\bibitem{Garzelli:2024uhe}
M.~V. Garzelli, G.~Limatola, S.~O. Moch, M.~Steinhauser, and O.~Zenaiev,
  ``{Updated predictions for toponium production at the LHC},'' {\em Phys.
  Lett. B} {\bf 866} (2025) 139532,
  \href{http://xxx.lanl.gov/abs/2412.16685}{{\tt 2412.16685}}.

\bibitem{Garzelli:2026ctb}
M.~V. Garzelli, G.~Limatola, S.-O. Moch, M.~Steinhauser, and O.~Zenaiev,
  ``{Threshold Top-Quark Pair-Production: Cross Sections and Key
  Uncertainties},'' \href{http://xxx.lanl.gov/abs/2604.09485}{{\tt
  2604.09485}}.

\bibitem{Sjostrand:2025qez}
T.~Sj{\"o}strand, ``{On the threshold behaviour of heavy top production},''
  {\em PoS} {\bf QCDEX2025} (2026) 002,
  \href{http://xxx.lanl.gov/abs/2510.04590}{{\tt 2510.04590}}.

\bibitem{Bierlich:2022pfr}
C.~Bierlich {\em et.~al.}, ``{A comprehensive guide to the physics and usage of
  PYTHIA 8.3},'' {\em SciPost Phys. Codeb.} {\bf 2022} (2022) 8,
  \href{http://xxx.lanl.gov/abs/2203.11601}{{\tt 2203.11601}}.

\bibitem{Breit:1936zzb}
G.~Breit and E.~Wigner, ``{Capture of Slow Neutrons},'' {\em Phys. Rev.} {\bf
  49} (1936) 519--531.

\bibitem{Combridge:1978kx}
B.~L. Combridge, ``{Associated Production of Heavy Flavor States in p p and
  anti-p p Interactions: Some QCD Estimates},'' {\em Nucl. Phys. B} {\bf 151}
  (1979) 429--456.

\bibitem{Ball:2013hta}
{\bf NNPDF} Collaboration, R.~D. Ball, V.~Bertone, S.~Carrazza, L.~Del~Debbio,
  S.~Forte, A.~Guffanti, N.~P. Hartland, and J.~Rojo, ``{Parton distributions
  with QED corrections},'' {\em Nucl. Phys. B} {\bf 877} (2013) 290--320,
  \href{http://xxx.lanl.gov/abs/1308.0598}{{\tt 1308.0598}}.

\bibitem{Skands:2014pea}
P.~Skands, S.~Carrazza, and J.~Rojo, ``{Tuning PYTHIA 8.1: the Monash 2013
  Tune},'' {\em Eur. Phys. J. C} {\bf 74} (2014), no.~8 3024,
  \href{http://xxx.lanl.gov/abs/1404.5630}{{\tt 1404.5630}}.

\bibitem{Sommerfeld:1931qaf}
A.~Sommerfeld, ``{{\"U}ber die Beugung und Bremsung der Elektronen},'' {\em
  Annalen Phys.} {\bf 403} (1931), no.~3 257--330.

\bibitem{Gamow:1928zz}
G.~Gamow, ``{Zur Quantentheorie des Atomkernes},'' {\em Z. Phys.} {\bf 51}
  (1928) 204--212.

\bibitem{Sakharov:1948plh}
A.~D. Sakharov, ``{Interaction of an Electron and Positron in Pair
  Production},'' {\em Zh. Eksp. Teor. Fiz.} {\bf 18} (1948) 631--635.

\bibitem{Barbieri:1973lza}
R.~Barbieri, P.~Christillin, and E.~Remiddi, ``{Vacuum Polarization and
  Positronium-Ground-State Splitting},'' {\em Phys. Rev. A} {\bf 8} (1973),
  no.~5 2266--2271.

\bibitem{Fadin:1995fp}
V.~S. Fadin, V.~A. Khoze, A.~D. Martin, and W.~J. Stirling, ``{Higher order
  Coulomb corrections to the threshold $e^{+} e^{-} \to W^{+} W^{-}$
  cross-section},'' {\em Phys. Lett. B} {\bf 363} (1995) 112--117,
  \href{http://xxx.lanl.gov/abs/hep-ph/9507422}{{\tt hep-ph/9507422}}.

\bibitem{Sjostrand:1993yb}
T.~Sj{\"o}strand, ``{High-energy physics event generation with PYTHIA 5.7 and
  JETSET 7.4},'' {\em Comput. Phys. Commun.} {\bf 82} (1994) 74--90.

\bibitem{PDF4LHCWorkingGroup:2022cjn}
{\bf PDF4LHC Working Group} Collaboration, R.~D. Ball {\em et.~al.}, ``{The
  PDF4LHC21 combination of global PDF fits for the LHC Run III},'' {\em J.
  Phys. G} {\bf 49} (2022), no.~8 080501,
  \href{http://xxx.lanl.gov/abs/2203.05506}{{\tt 2203.05506}}.

\bibitem{Mertig:1990an}
R.~Mertig, M.~Bohm, and A.~Denner, ``{FEYN CALC: Computer algebraic calculation
  of Feynman amplitudes},'' {\em Comput. Phys. Commun.} {\bf 64} (1991)
  345--359.

\bibitem{Shtabovenko:2020gxv}
V.~Shtabovenko, R.~Mertig, and F.~Orellana, ``{FeynCalc 9.3: New features and
  improvements},'' {\em Comput. Phys. Commun.} {\bf 256} (2020) 107478,
  \href{http://xxx.lanl.gov/abs/2001.04407}{{\tt 2001.04407}}.

\bibitem{Ruijl:2017dtg}
B.~Ruijl, T.~Ueda, and J.~Vermaseren, ``{FORM version 4.2},''
  \href{http://xxx.lanl.gov/abs/1707.06453}{{\tt 1707.06453}}.

\bibitem{Beneke:2024sfa}
M.~Beneke and Y.~Kiyo, ``{Third-order correction to top-quark pair production
  near threshold II. Potential contributions},'' {\em JHEP} {\bf 07} (2025)
  274, \href{http://xxx.lanl.gov/abs/2409.05960}{{\tt 2409.05960}}.

\bibitem{Nason:2025hix}
P.~Nason, E.~Re, and L.~Rottoli, ``{Spin correlations in $ t\overline{t} $
  production and decay at the LHC in QCD perturbation theory},'' {\em JHEP}
  {\bf 10} (2025) 149, \href{http://xxx.lanl.gov/abs/2505.00096}{{\tt
  2505.00096}}.

\end{thebibliography}\endgroup

\end{document}